\documentclass[letterpaper,twocolumn,10pt]{article}
\usepackage{usenix2020-09}
\usepackage{amssymb} 
\usepackage{tikz}
\usepackage{amsmath}
\usepackage{url}
\usepackage{multirow}
\usepackage[linesnumbered,ruled,vlined]{algorithm2e}

\usepackage{mdframed}
\mdfdefinestyle{dashedbox}{
    linecolor=black,
    linewidth=1pt,
    innertopmargin=5pt,
    innerbottommargin=5pt,
    innerrightmargin=5pt,
    innerleftmargin=5pt
}

\usepackage{balance}

\newcommand{\sysname}{\textsc{DeepServe}}
\newcommand{\flowserve}{\textsc{FlowServe}}
\newcommand{\autoscaler}{\textsc{AutoScaler}}
\newcommand{\cloudmatrix}{\textsc{CloudMatrix384}}

\newcommand{\ys}[1]{{{\color{red}#1}}{}}

\usepackage{tikz}

\begin{document}
\pagestyle{empty} % For camera ready version

\date{}

\title{\sysname: Serverless Large Language Model Serving at Scale}

\author{
{\rm Junhao Hu}$^{23*}$, 
{\rm Jiang Xu}$^{1}$, 
{\rm Zhixia Liu}$^{1}$, 
{\rm Yulong He}$^{1}$, 
{\rm Yuetao Chen}$^{1}$, 
{\rm Hao Xu}$^{1}$, 
{\rm Jiang Liu}$^{1}$,
{\rm Jie Meng}$^{1}$,  \\
{\rm Baoquan Zhang}$^{1}$,
{\rm Shining Wan}$^{1}$, 
{\rm Gengyuan Dan}$^{1}$, 
{\rm Zhiyu Dong}$^{1}$, 
{\rm Zhihao Ren}$^{1}$, 
{\rm Changhong Liu}$^{1}$, \\
{\rm Tao Xie}$^{32}$,
{\rm Dayun Lin}$^{1}$, 
{\rm Qin Zhang}$^{1}$, 
{\rm Yue Yu}$^{1}$, 
{\rm Hao Feng}$^{1}$, 
{\rm Xusheng Chen}$^{1\dag}$, 
{\rm Yizhou Shan}$^{1\dag}$
\\
\\
$^{1}$Huawei Cloud  
$^{2}$SCS, Peking University
$^{3}$Key Lab of HCST (PKU), MOE
}

\maketitle

\renewcommand{\thefootnote}{\fnsymbol{footnote}}
\footnotetext[1]{This work was completed during his internship at Huawei Cloud.}
\footnotetext[2]{Co-corresponding authors.}
\renewcommand{\thefootnote}{\arabic{footnote}}

\begin{abstract}

In this paper, we propose \sysname, a scalable and serverless AI platform designed to efficiently serve large language models (LLMs) at scale in cloud environments. \sysname\ addresses key challenges such as resource allocation, serving efficiency, and cold start latencies through four main design components.
First, \sysname\ uses a simple serverless abstraction called the request-job-task model, which helps manage diverse AI workloads across post-training and model-serving tasks.
Second, \sysname\ integrates an in-house serving engine named \flowserve\ using a microkernel-inspired design, NPU-centric execution, and SPMD-based parallelism to optimize LLM serving. 
Third, \sysname\ includes novel scheduling policies tailored for a configuration with both PD-disaggregated and PD-colocated instances. Fourth, \sysname\ includes optimizations such as pre-warmed pods, DRAM pre-loading, and NPU-fork, which allow \sysname\ to scale up to 64 instances in seconds.
\sysname\ has been in production for over a year, operating on a large Ascend NPU cluster and providing industry-standard APIs for fine-tuning, agent serving, and model serving to our customers.
\end{abstract}
\section{Introduction}

The rapid rise of generative AI, exemplified by the success of platforms such as ChatGPT, has transformed the landscape of AI and increased the demand for scalable systems capable of serving large language models (LLMs).
Model-as-a-Service (MaaS) platforms, such as OpenAI's offerings, enable millions of users to access powerful AI capabilities through a simple API interface. Consequently, LLM serving has become one of the most crucial workloads in modern data centers.

Ensuring such services' performance, efficiency, and cost-effectiveness is complex, and we identify the following three challenges to achieve optimal performance and resource utilization while ensuring Service Level Objectives (SLOs) for multi-tenant users. 
First, AI workloads vary significantly in duration, from fine-tuning that can last hours or even days, to agent serving and LLM serving, which typically range from seconds to minutes. This variation makes it difficult to dynamically allocate shared resources without either underutilizing or overloading the system.
Second, as LLM serving becomes increasingly distributed and stateful, managing resource allocation, synchronization, and fault tolerance across multiple instances is more complex.
A single inference request may span multiple distributed instances and involve cached states, creating additional challenges in ensuring efficient operation.
Third, the variability in LLM serving demands leads to fluctuating resource needs, further complicating resource optimization and the handling of cold-start latencies~\cite{fu2024serverlessllm}.

To address the preceding challenges, in this paper, we propose \sysname, Huawei Cloud's fully-hosted and serverless platform, offering industry-standard fine-tuning, agent serving, and model-serving APIs~\cite{openai-url}. \sysname\ has been running for over a year atop a large Ascend NPU cluster~\cite{ascend-npu-hotchips19}. We report how \sysname\ addresses the preceding challenges using four major design components: serverless abstraction and infrastructure, serving engine, scheduling algorithms, and scaling optimizations.

\textbf{Serverless Infrastructure.} \sysname\ introduces a developer-facing serverless abstraction called the request-job-task model. In this model, a \textit{request} is an external trigger (e.g., a user-sent HTTP call). A \textit{job} (or multiple \textit{jobs}) matching the request type (e.g., chat, fine-tuning) handles the request. A task is a fine-grained operation within a job (e.g., prefill task, decode task).
Users send HTTP \textit{requests} that trigger internal \textit{jobs}, which are broken down into smaller \textit{tasks}.
This request-job-task abstraction allows for the dynamic scaling of workloads and efficient resource sharing across post-training and model-serving tasks.
\sysname's architecture, as shown in Figure~\ref{fig-system-overview}, is built around this abstraction and consists of three core components: Job Executors (JEs), Task Executors (TEs), and a cluster manager.
JEs handle incoming requests and decompose them into manageable tasks that are then distributed to TEs for execution. The cluster manager ensures the health and scaling of the job and task executors.
\sysname\ deploys JEs and TEs dedicated to post-training and serving. We focus on serving in this paper.

\textbf{Serving Engine}.
In \sysname, we design an efficient serving engine called \flowserve, built on three fundamental principles.
The first is a microkernel-inspired design, which divides system functionality into modular components that can scale independently. This separation ensures that different system parts can evolve and operate asynchronously.
The second principle is NPU-centric execution, which aims to keep the NPU busy to minimize delays caused by other resources such as CPUs, DRAM, or storage.
Finally, \flowserve\ adopts a Single-Program-Multiple-Data (SPMD)-based design, enabling efficient parallel processing and scaling across multiple NPUs.

The \flowserve\ engine has six core functionalities:
tokenization,
model execution,
scheduler,
memory management,
caching management, and networking management.
The tokenizer operates independently.
Adhering to the SPMD design principle,
\flowserve\ follows a master-executor architecture: the master oversees scheduling, caching, and networking decisions, while per-NPU executors carry out these decisions on their respective NPUs.
\flowserve's scheduler is centralized at the master module, using both asynchronous KV cache prefetch and asynchronous execution to keep an NPU busy.
\flowserve\ has a Relational Tensor Cache (RTC) module to manage the relationship between tensors (primarily on the KV cache) and a distributed flow (DistFlow) module to transfer tensors across tiered storage within a single engine or across engines in a peer-to-peer manner.

\textbf{Scheduling Algorithm.}
\sysname\ presents three designs to tackle the challenges introduced by prefix caching and disaggregation.
First, we implement a locality-aware algorithm to maximize KV cache reuse, a design shared by previous work~\cite{srivatsa2024preble, sglang, hu2024memserve}.
Second, we extensively compare PD-disaggregated and PD-colocated TEs in a controlled environment with varying configurations.
Based on the profiling results, we develop a PD-aware scheduling policy that accounts for the dynamics of online serving and decode-length uncertainty.
Finally, we propose a combined scheduling algorithm that integrates load-aware, locality-aware, and PD-aware strategies.

\textbf{Scaling Optimization.}
\sysname\ achieves fast scaling by quickly adjusting to fluctuating workloads. 
It employs multiple key techniques, including reserving pre-warmed Pods and TEs, pre-loading models into DRAM, and leveraging high-speed NPU-to-NPU links for efficient model loading.
Combined with optimizations such as parallel initialization and predictive model pre-loading, these techniques significantly reduce initialization time.
\sysname\ can scale up to 64 instances in parallel within seconds.

This paper makes the following main contributions:
\begin{itemize}
\item Design of \sysname, a large-scale serverless AI platform for LLM serving (\S\ref{sec:design-overview}).

\item Design of \flowserve, an efficient and modular serving engine architecture (\S\ref{sec:flowserve}).

\item Study of scheduling techniques for PD-disaggregated and PD-colocated setups (\S\ref{sec-dist-sched}).

\item Detailed description of optimization techniques for fast scaling in LLM serving (\S\ref{sec-scaling}).
\end{itemize}
\section{Background}

We describe LLM and Ascend NPU chips in this section.

\textbf{LLM.}
LLM inference consists of two main stages: the \emph{prefill stage} and the \emph{decode stage}. In the prefill stage, the model processes the full input prompt $(x_1, x_2, \dots, x_n)$, computes Key-Value (KV) vectors for each token, and stores them in the KV cache~\cite{vllm-sosp23, hu2025epic, hu2025raas}. The model then generates the next token $x_{n+1}$ to initiate the decode stage. This stage is compute-bound and benefits from high parallelism.
In the decode stage, the model iteratively generates one token at a time. For each step, it computes and appends the corresponding KV vectors to the KV cache. Token generation continues until a stopping condition is met. This stage is memory-bound, dominated by cache lookup and memory access rather than computation.

\textbf{Ascend NPU and SuperPod.}
Our system runs on Huawei's Ascend Neural Processing Unit (NPU) AI chips~\cite{ascend-npu-hotchips19}. The Ascend 910B provides 400~TFLOPS of FP16 compute and 64\,GB of HBM. The newer Ascend 910C consists of two dies, each offering 400~TFLOPS and 64\,GB of HBM.
We currently support two generations of NPU clusters (see Figure~\ref{fig-system-overview}).
The first generation uses a scaled-out architecture. Each server contains 8 Ascend 910B chips, and servers are connected via a 200\,Gbps RoCE network.
The second generation adopts a scaled-up architecture, recently released as the \cloudmatrix\ SuperPod~\cite{url-cloudmatrix}.
It comprises 48 servers and 384 Ascend 910C chips, all interconnected via a high-bandwidth network around 200\,GB/s (unidirectional). All CPUs and NPUs in the SuperPod share a unified memory address space, enabling access to any chip's DRAM or HBM.

\if 0
In an NPU cluster, two types of communication links are available for NPU-to-NPU communication.
%
The first is a scaled-up network (e.g., HCCS) that provides higher bandwidth and lower latency.
The second is a scaled-out network (e.g., RoCE) that supports larger-scale deployments, offering lower bandwidth but enabling the connection of thousands of nodes.
%
To facilitate efficient communication across the NPU cluster, Huawei provides the Huawei Collective Communication Library (HCCL), which includes collective APIs such as \textit{all\_reduce} and \textit{broadcast}, as well as peer-to-peer APIs such as \textit{send} and \textit{recv}.
The hardware trend is that the scaled-up network domain is increasingly larger, enabling us to move from regular servers to so-called SuperPod and eventually fully disaggregated data centers~\cite{fully-disagg-apsys22}.

\textbf{SuperPod.}
%
CloudMatrix384~\cite{url-cloudmatrix}.
\fi 
\section{\sysname: A Serverless AI Platform}
\label{sec:design-overview}

{
\begin{figure*}[t]
\begin{center}
\centerline{\includegraphics[width=\textwidth]{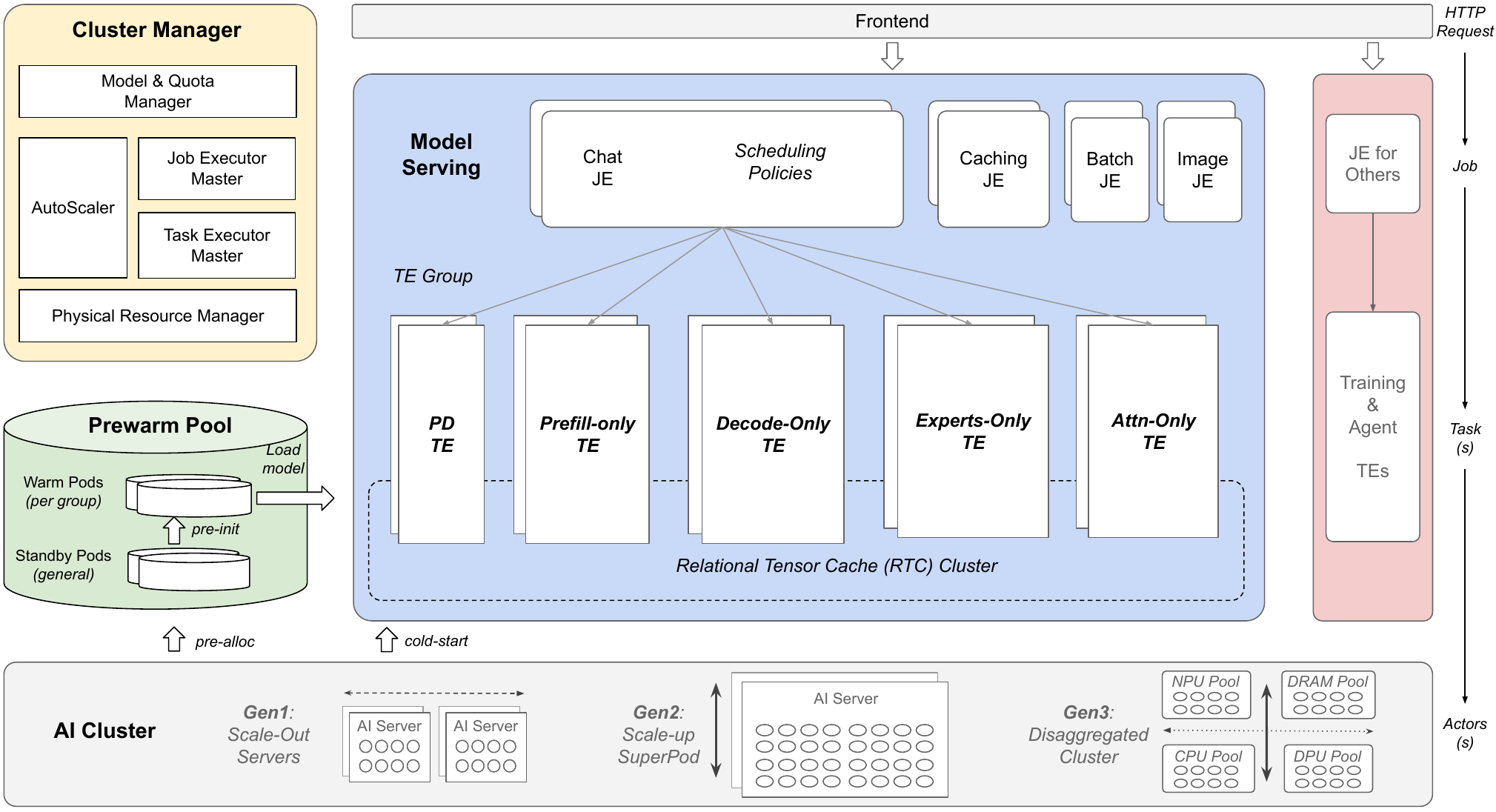}}
\caption{\textbf{\sysname\ Overall Architecture}.
(a) User requests are routed through the frontend and dispatched to the appropriate JEs.
(b) Model serving includes various types of JEs, such as those for chat completion and batch inference.
(c) Each model-serving JE independently runs distributed scheduling algorithms.
(d) Model-serving TEs in an RTC group can exchange tensors using DistFlow APIs (see \S\ref{sec:flowserve-distflow}).
(e) The cluster manager is a centralized, high-availability module.
(f) We omit post-training and agent serving for brevity.
(g) We plan to run three generations of the NPU cluster. We currently support Gen1 and Gen2.
}
\label{fig-system-overview}
\end{center}
\end{figure*}
}

\sysname\ is Huawei Cloud's fully-hosted and serverless platform for running emerging generative AI workloads, offering industry-standard fine-tuning, agent serving, and model-serving APIs~\cite{openai-url}. \sysname\ has been running for over a year atop a large Ascend NPU cluster. As a cloud platform, our goal is to maximize AI cluster \textit{performance} and \textit{utilization} while ensuring SLO guarantees for multi-tenant users. However, achieving this goal with emerging AI workloads faces the following three key challenges.
\begin{itemize}
%\setlength\itemsep{0em} % Reduce the space between items
    % ---> Overall design
    \item Challenge 1: AI workloads have varying durations, posing challenges for efficient resource sharing. For example, 
    fine-tuning can take hours to days, while agent serving and LLM serving typically take seconds to minutes.
    The challenge lies in dynamically allocating shared resources among these workloads without underutilizing or overloading the system.

    % ---> FlowServe, Dist Serving and Dist Scheduling
    \item Challenge 2: LLM serving is becoming more distributed and stateful. A single inference request may span multiple distributed instances~\cite{patel2023splitwise,tetriserve-arxiv24,zhong2024distserve} and use cached states~\cite{sglang,hu2024memserve,hu2025epic}, making it challenging to efficiently manage resource allocation, synchronization, and fault tolerance across the system.

    % ---> Fast Auto Scaling
    \item Challenge 3: LLM serving is highly variable, leading to fluctuating resource demands~\cite{blizscale-arxiv24,fu2024serverlessllm}. This fluctuation introduces challenges in optimizing resource utilization and handling cold-start latencies.
\end{itemize}

To address these challenges, \sysname\ (Figure~\ref{fig-system-overview}) integrates the following four key designs.

\textbf{Serverless Abstraction}.
\sysname\ introduces a developer-facing serverless abstraction called the request-job-task model. In this model, a \textit{request} is an external trigger (e.g., a user-sent HTTP call). A \textit{job} (or multiple \textit{jobs}) matching the request type (e.g., chat, fine-tuning) handles the request. A task is a fine-grained operation within a job (e.g., prefill task, decode task).
Users interact with \sysname\ by sending HTTP requests, each triggering one or more internal jobs. A job, in turn, may spawn multiple tasks.
For example, a fine-tuning request triggers multiple jobs, such as preprocessing, training, and evaluation. A chat request triggers a single serving job. In the chat case, if executed on a PD-colocated engine~\cite{vllm-sosp23}, the job generates one task. If executed in a prefill-decode-disaggregated setup~\cite{patel2023splitwise}, the job generates two tasks: one for the prefill instance and the other for the decode instance.
This abstraction allows \sysname\ to
(i) scale AI workloads over shared infrastructure,
(ii) consolidate post-training and serving on the same cluster, and
(iii) simplify distributed LLM serving.

More specifically, \sysname's serverless abstraction consists of three core components: job executors, task executors, and a cluster manager.
The Job Executor (JE) processes incoming requests, decomposes them into tasks, and assigns tasks to available task executors for execution.
The task executor (TE) is responsible for executing the tasks.
\sysname\ deploys JEs and TEs dedicated to post-training, agent serving, and model serving.
The cluster manager is a highly available system that oversees and scales all JEs and TEs.
It includes centralized master modules for both JEs and TEs,
each monitoring their health.
Due to space limits, this paper focuses on the model-serving aspect.

\textbf{Efficient LLM Serving.}
We design \flowserve\ (\S\ref{sec:flowserve}) as an efficient serving engine used by each model-serving TE.
To minimize interference between the prefill and decode stages of LLMs, \flowserve\ adopts a PD-disaggregated serving paradigm for both dense and sparse models~\cite{zhong2024distserve, hu2024memserve}.

\textbf{Scheduling.} To support increasingly stateful and disaggregated serving, we propose distributed scheduling algorithms (\S\ref{sec-dist-sched}) that run on model-serving JEs.
Model-serving TEs running the same model and serving mode (e.g., prefill-only, decode-only) are organized into a TE group. JEs dispatch requests to appropriate TEs within these groups based on our scheduling algorithms.

\textbf{Fast Scaling.}
\sysname\ uses fast scaling (\S\ref{sec-scaling}) to quickly adjust to changing online workloads. Key techniques include reserving pre-warmed Pods and TEs, pre-loading models into DRAM, and using high-speed NPU-to-NPU links for faster model loading. These techniques greatly reduce code-start time and allow \sysname\ to scale up to 64 instances in parallel within seconds.

In the rest of the paper, we discuss \textbf{serving} in \S\ref{sec:flowserve}, \textbf{scheduling} in \S\ref{sec-dist-sched}, and \textbf{scaling} in \S\ref{sec-scaling}.
\section{\flowserve: An Efficient Serving Engine}
\label{sec:flowserve}

This section presents \flowserve, our in-house serving engine for LLMs, Large Multimodal Models (LMM), and embedding models.

{
\begin{figure*}[t]
\begin{center}
\centerline{\includegraphics[width=\textwidth]{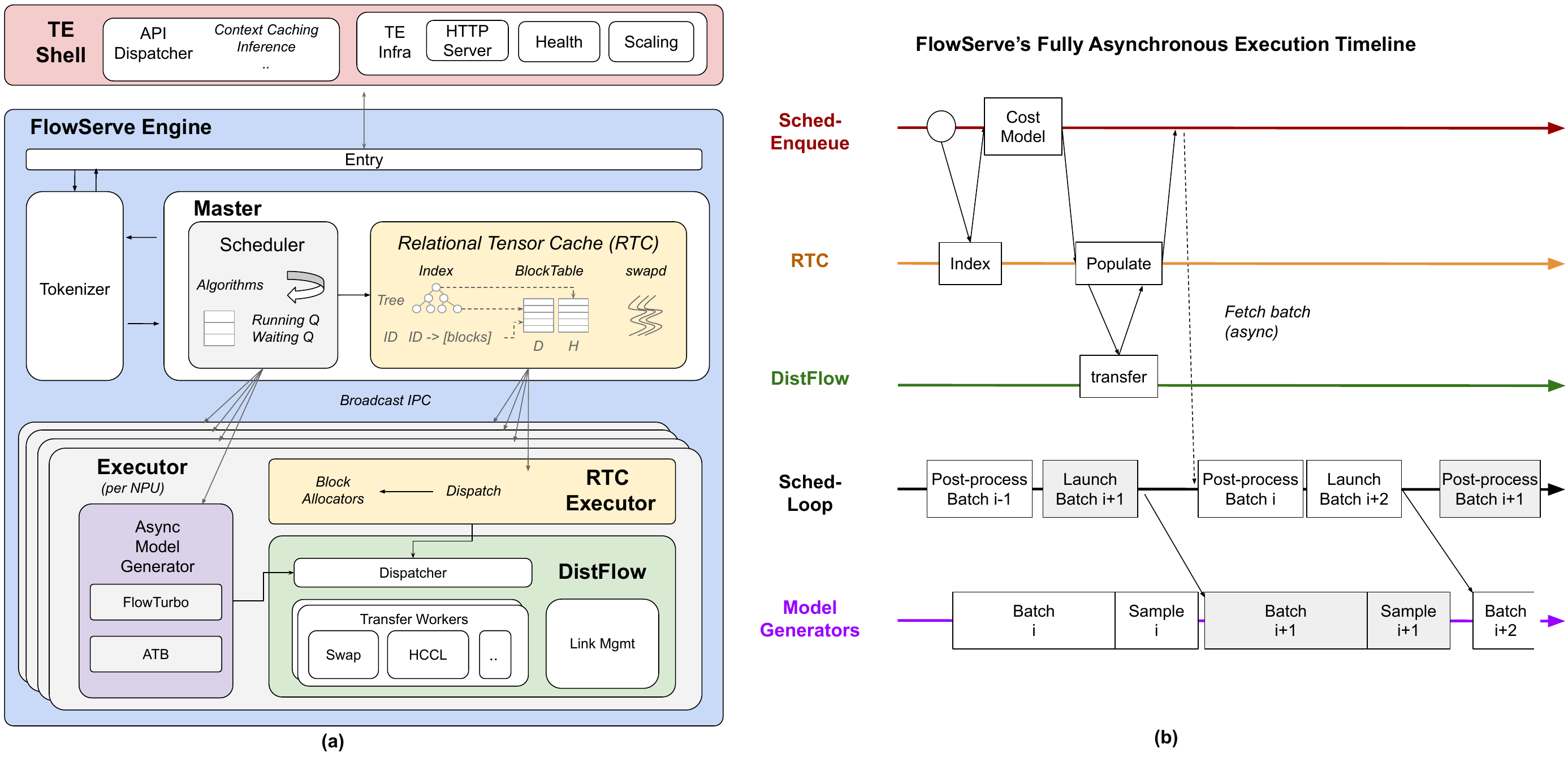}}
\caption{\textbf{\flowserve\ Architecture and Asynchronous Execution Timeline.}
(a) The \flowserve\ engine can be deployed without the TE-shell.
(b) The \flowserve's model generator supports multiple backends. FlowTurbo is a torch-based dynamic graph. Ascend-Transformer-Boost (ATB) is a C++-based static graph~\cite{mindie-atb}.
}
\label{fig-flowserve}
\end{center}
\end{figure*}
}

% (c) The \flowserve\ engine is highly scalable, supporting deployments ranging from a single chip (TP=1) to clusters with tens of servers and hundreds of chips.\textbf{}

\subsection{Overview}

We build \flowserve\ with the following three main goals\footnote{These goals were established in late 2023 and have guided the design ever since.}. 
First, we want to create a high-throughput serving engine that works as a single TE. 
Second, we aim for fast iteration, given that serving is a fast-moving field. 
Third, we want \flowserve\ to be Ascend-NPU-native, able to work with different generations of the Ascend cluster---from regular scaled-out servers to scaled-up SuperPod, and eventually fully disaggregated data centers~\cite{fully-disagg-apsys22}.

To achieve these goals, we build \flowserve\ based on three guiding principles:
\begin{itemize}
    \item \textbf{Microkernel-Inspired Design.} We apply the core principles of microkernel architecture by decoupling serving-engine functionalities into modular components. This modularity enables independent scaling, independent evolution, and asynchronous operations, thereby enhancing throughput.
    
     \item \textbf{NPU-Centric Execution.} We aim to keep NPUs busy all the time, reducing delays caused by waiting for other resources, such as CPU, DRAM, storage, and network.
    
    \item \textbf{SPMD-based Design.} We use the Single-Program-Multiple-Data (SPMD~\cite{pathways-mlsys22}) design to enable efficient parallel processing and scaling across multiple NPUs, a design shared by vLLM~\cite{vllm-sosp23}.
\end{itemize}

Figure~\ref{fig-flowserve} presents \flowserve's overall architecture.
Each model-serving TE has a \textit{TE-shell} and a \flowserve\ engine.
The \textit{TE-shell} is an infrastructure module consisting of predefined modules such as scaling, health reporting, and a few customized modules that serve as context caching handlers.

The \flowserve\ engine has six core functionalities:
tokenization (with a tokenizer),
model execution,
scheduler,
memory management,
caching management,
and networking management.
The tokenizer is an independent module that can scale on its own.
Adhering to the SPMD design principle,
\flowserve\ follows a master-executor architecture:
the master oversees scheduling, caching, networking decisions,
while per-NPU executors carry out these decisions on their respective NPUs.
We map the remaining functionalities, excluding the tokenization, onto this master-executor design as follows:
\begin{itemize}

    \item \textbf{Scheduling and Model Execution (\S\ref{sec:flowserve-sched}).} The master scheduler assigns the next batch, and each executor runs the model's forward pass on this batch. The master broadcasts requests to all executors when initiating a batch.
    
    \item \textbf{Caching and Memory Management (\S\ref{sec:flowserve-rtc}).}
    We build a Relational Tensor Cache (RTC) to unify the management of both caching and memory allocation.
    The master maintains indexing structures such as a prefix tree~\cite{sglang}, while each executor handles memory allocation.
    
    \item \textbf{Networking Management (\S\ref{sec:flowserve-distflow}).}
    We build Distributed Flow (DistFlow) to transfer tensors between model-serving TEs.
    DistFlow runs on each executor, providing memory-transfer APIs for both the model generator and scheduler, supporting multiple transfer backends and Ascend cluster generations.
\end{itemize}

We implement \flowserve\ primarily in Python, with RTC and DistFlow implemented in C++.
Although having design principles is important, translating them into an efficient implementation is a significant challenge. 
Figure~\ref{fig-flowserve-perf} shows the decoding performance of \flowserve\ in three versions over a period of three months.
From v1 to v2, we introduced asynchronous scheduling and IPC optimization, which resulted in more than 2x improvements when the TPOT SLO was set to 50ms.
From v2 to v3, our optimizations focused on data structures, sampling, and so forth; these optimizations resulted in roughly 20\% improvement.

\if 0
\ys{
1. centralized scheduler v.s. scheduler in each executor (sglang)
2. PP adopts a centralized scheduler. vLLM, each PP stage has a scheduler
}
\fi

\subsection{Scheduling}
\label{sec:flowserve-sched}

The scheduler is centralized in the master module. We choose this approach over the distributed schedulers (used by SGLang~\cite{sglang}) due to its simplicity, albeit at the cost of frequent Inter-Process Communications (IPCs).
The scheduler supports asynchronous KV cache prefetching enabled by RTC and DistFlow (described below)
and asynchronous execution, a design shared by vLLM~\cite{vllm-sosp23}, SGLang~\cite{sglang}, and NanoFlow~\cite{zhu2024nanoflow}.
We now describe the execution timeline in the right part of Figure~\ref{fig-flowserve}.

\textbf{Asynchronous KV Cache Prefetch.}
A sched-enqueue thread handles tokenized requests from the tokenizer with the following three steps.
First, the thread calls the RTC \texttt{match} API to check for any preserved KV cache.
The returned info tells whether a prompt prefix or ID has cached KV and the cached KV's location (e.g., in the NPU or swapped to tiered storage).
Second, if there is cached KV but a portion of it is not in the NPU, the scheduler runs a cost model to decide whether reusing the cache is beneficial.
If the cost model suggests that reusing the cache can improve performance, the scheduler calls the \texttt{populate} API to request RTC to fetch the cache into the NPU.
This step is done asynchronously. RTC will call DistFlow to read KV from tiered storage or other TEs.
Once RTC completes the cache population, it notifies the sched-enqueue thread, which marks the request as ready.
The sched-loop will pick it up during the next scheduling cycle.

\textbf{Asynchronous Execution.}
The asynchronous execution design aligns with our NPU-centric principle of fully utilizing the NPU.
It is similar to the new zero-overhead scheduler introduced in SGLang-v0.4.0~\cite{sglang}.
The key observation behind this design is that scheduling decisions do not depend on the actual token IDs generated by the model, but rather on the number of tokens to be processed in each run.
In typical decoding scenarios, where one token is generated per sequence per decoding step, the scheduler can predict the required resources for the next run before the current one completes. Doing so allows the scheduler to operate in a separate thread, preparing the necessary inputs for the model generator beforehand.
By running the scheduler in parallel with the model execution, we eliminate unnecessary CPU wait time, ensuring that the NPU remains busy.

\if 0
\textbf{Policy.}
Our scheduler incorporates priority-aware scheduling. For sequences in the waiting queue, it prioritizes those with the highest priority, and for sequences with equal priority, it adopts a first-come, first-served (FCFS) approach. When key-value (KV) cache capacity is exhausted and no additional memory blocks can be allocated for the next run, the scheduler preempts the sequence with the lowest priority. Preemption is implemented through recomputation or swapping. Due to the scheduler's asynchronous design, swapping occurs immediately after the preemption decision, in parallel with the current forward pass, thereby minimizing interference with model execution.
\fi

\textbf{Pipeline Parallelism (PP).}
We optimize our scheduler for PP by running a centralized scheduler at the first stage of PP; other stages accept only requests from previous stages.
This design enables \flowserve\ to manage all micro-batches in a unified way. This approach has two main benefits.
(1) Memory resources are managed in one place, making it easy to preempt sequences across micro-batches.
(2) With chunked prefill enabled, the scheduler distributes chunks across consecutive micro-batches, rather than sticking to just one micro-batch~\cite{sarathi-arvix23}. Doing so helps reduce Time-To-First-Token (TTFT) by at least 20\%.

%
%In short, extending \flowserve\ to support DP attention requires efficient scalability and networking management.

{
\begin{figure}[t]
\begin{center}
\centerline{\includegraphics[width=\linewidth]{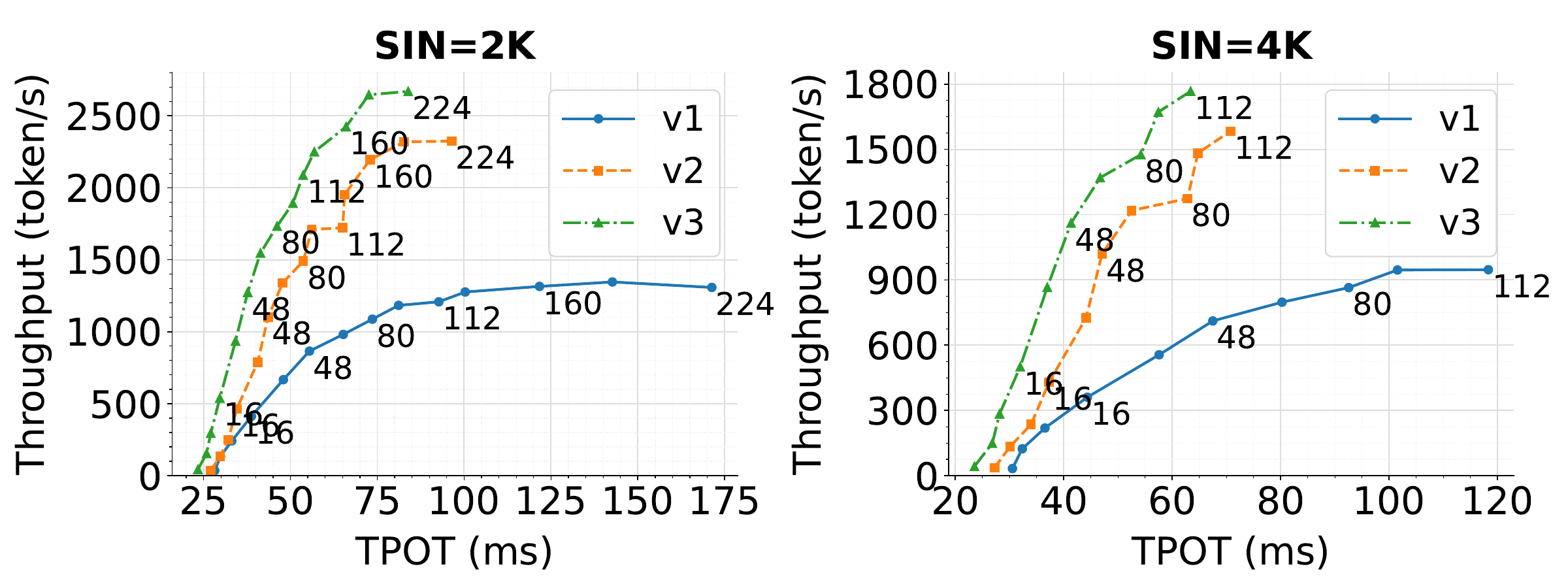}}
\caption{\textbf{\flowserve\ Offline Serving Perf.} We run a 34B model with TP=4. The left has a prefill length of 2K, and the right is 4K. We run 256 decoding steps and report the average TPOT and decoding throughput.}
\label{fig-flowserve-perf}
\end{center}
\end{figure}
}

\subsection{Caching}
\label{sec:flowserve-rtc}

\begin{table}
    \caption{The Core APIs of Relational Tensor Cache.}
    \vspace{0.1in}
    \footnotesize
    \centering
    \begin{tabular}{p{1in} | p{1.6in} }
    \hline
    \textbf{API} & \textbf{Description} \\
    \hline
    \hline
       \texttt{MatchByPrefixToken} &  Find preserved KV cache by tokens \\
    \hline
       \texttt{MatchByID} &  Find preserved KV cache by ID \\
    \hline
       \texttt{Populate} &  Fetch preserved KV cache into NPU\\
    \hline
       \texttt{QueryPopulate} &  Check populate status \\
    \hline
       \texttt{AllocBlocks} &  Alloc blocks for prefill \\
    \hline
       \texttt{AppendBlock} &  Alloc block for decode \\
    \hline
       \texttt{Copy} &  Copy blocks from NPU to DRAM \\
    \hline
       \texttt{Free} &  Free blocks \\
    \hline
    \end{tabular}
    \label{tbl-rtc-api}
\end{table}

We build a Relational Tensor Cache (RTC) to integrate prefix caching~\cite{sglang} and position-independent caching~\cite{yao2024cacheblend, hu2025epic} into \flowserve.
The core function of RTC is to manage the \textit{relationship} between tensors, primarily the KV cache.

\textbf{Abstraction.}
Table~\ref{tbl-rtc-api} lists the main APIs of RTC.
RTC provides two main sets of APIs.
The first set is for managing basic blocks, such as \texttt{AllocBlocks} and \texttt{AppendBlock}, which are used by prefill and decode requests, respectively.
The second set is for managing KV caching.
Notably, RTC includes various match APIs, each supporting different indexing mechanisms, such as prefix-token-based or explicit ID-based matching.
\flowserve's implicit caching uses the prefix-token-based mechanism, while the explicit ID-based mechanism is used by \sysname's explicit context caching endpoint.
RTC also offers a novel \texttt{populate} API.
When invoked, RTC fetches the specified KV cache in the given address range into local NPUs.
%Since this operation may involve moving data across tiered storage or distributed TEs, RTC also includes a query API to handle such requests.

\textbf{Design.}
RTC follows the master-executor architecture of \flowserve.
The master module manages the RTC module, making decisions about allocation and data movement, while each executor runs an RTC executor to perform these actions.
Internally, RTC includes a traditional block table, originally proposed by vLLM~\cite{vllm-sosp23}, for managing data blocks.
Additionally, RTC employs a hybrid indexing layer that combines radix-tree indexing~\cite{sglang} with ID-based indexing.
In our current design, each index node can point to data stored either in the NPU or in local DRAM. We are also adapting RTC for SuperPod to leverage its global shared memory.
Finally, the RTC master runs multiple background threads to handle tasks such as background swapping and prefetching.

\subsection{Networking}
\label{sec:flowserve-distflow}

Emerging distributed serving techniques---such as disaggregated prefill-and-decode~\cite{zhong2024distserve, patel2023splitwise, tetriserve-arxiv24} and disaggregated attention-and-experts~\cite{pan2024instinfer, attention-offload-qinghua-2024}---require new communication primitives that go beyond traditional collectives.
\texttt{DistFlow} is designed to meet this demand.
\texttt{DistFlow} specializes in peer-to-peer and many-to-many tensor transfers across heterogeneous memory. It supports transfers both within a single TE across tiered storage and between distributed TEs.

\textbf{Abstraction.}
\texttt{DistFlow} provides a simple but expressive interface. Its control-plane APIs include \texttt{LinkCluster} for establishing peer groups, and its primary data-plane API is \texttt{transfer(srcInfo, dstInfo)}. Users are required to specify exact source and destination memory buffers; \texttt{DistFlow} operates on raw memory addresses rather than higher-level block abstractions. In \flowserve, this API is used by the centralized scheduler, each executor's model generator, and the RTC executor.

\textbf{Design.}
\texttt{DistFlow} follows key principles from high-performance RPC systems~\cite{erpc-nsdi19, kalia2016fasst, tsai2017lite, monga2021birds}. It employs scalable threading models to avoid synchronization bottlenecks and supports multiple backends to adapt to heterogeneous network fabrics.
In scaled-out Ascend clusters, \texttt{DistFlow} uses the Huawei Collective Communication Library (HCCL) peer-to-peer API as its transfer backend. In scaled-up Ascend SuperPods, where CPUs and NPUs share a unified memory space, it leverages NPU memory copy primitives for direct, high-bandwidth tensor transfers.

%
%We believe the system most similar to DistFlow is Mooncake's Transfer Engine~\cite{qin2024mooncake}.

\subsection{Disaggregated Serving}

{
\begin{figure}[t]
\begin{center}
\centerline{\includegraphics[width=\linewidth]{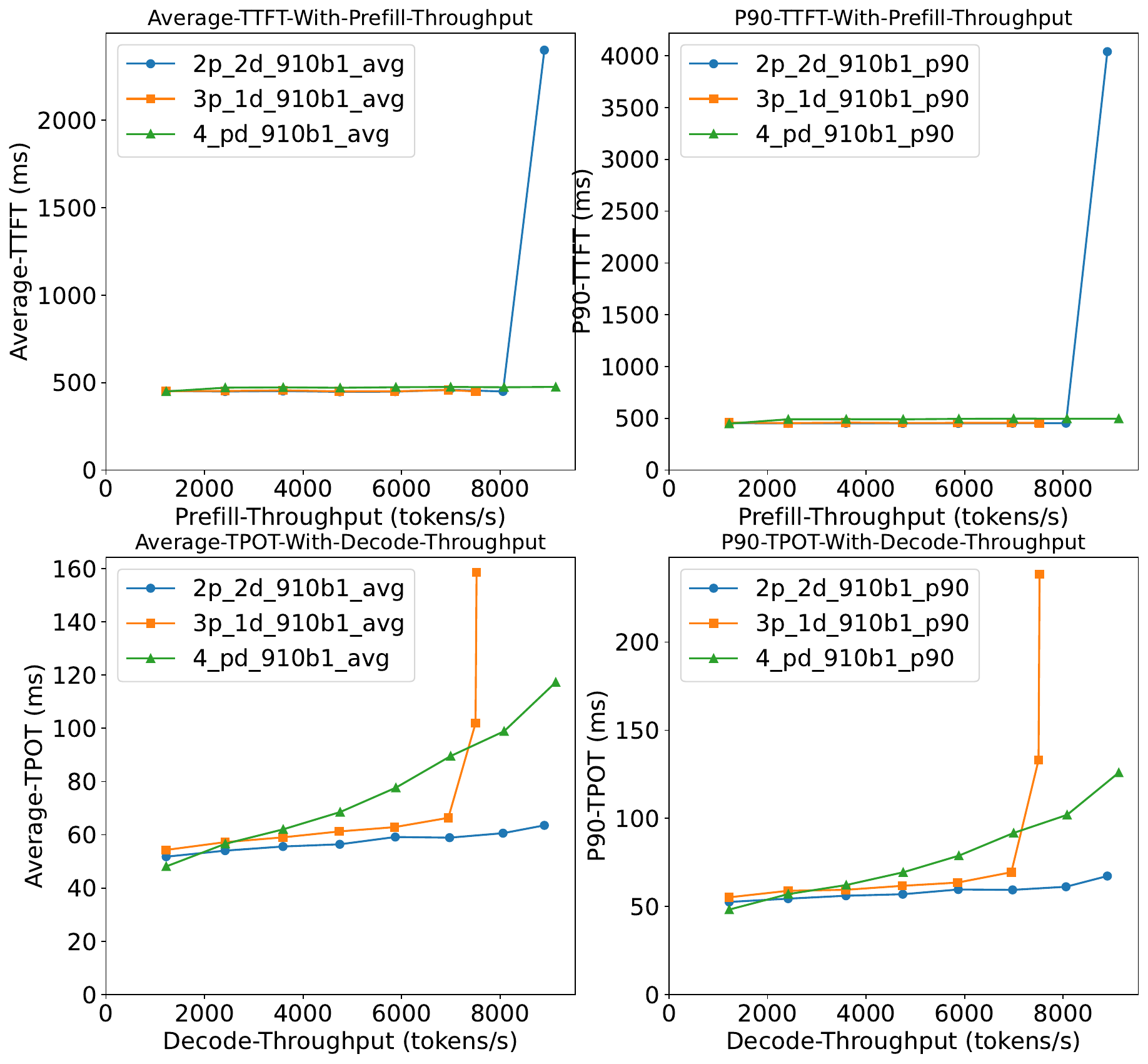}}
\caption{\textbf{\flowserve\ Online Serving Perf.} We run a 34B model with TP=4 using an internal trace (roughly 2K input with 200 output). We test three setups: (1) PD-disaggregated with two prefill and two decode, (2) PD-disaggregated with two prefill and one decode, and (3) four PD-colocated. We vary RPS from 0.2 to 1.2 in a step of 0.2.}
\label{fig-pd-study}
\end{center}
\end{figure}
}

Serving is increasingly disaggregated.
In \sysname, we define two levels of disaggregation:
\begin{itemize}
    \item \textbf{Task-level}.
    This level refers to disaggregating prefill and decode into separate TEs.
    Our implementation is similar to previous work~\cite{zhong2024distserve,patel2023splitwise,tetriserve-arxiv24}:
    \flowserve\ in the prefill TE sends prefilled KV cache to decoding TE either by-req or by-layer using DistFlow's transfer API.
    Figure~\ref{fig-pd-study} shows an online serving test comparing PD-disaggregated and PD-colocated using an internal trace. Disaggregation greatly improves throughput under certain SLOs and lowers TPOT with the same throughput.
    
    \item \textbf{Operator-level}.
    This level refers to disaggregating attention and experts into separate TEs.
    We take inspiration from pioneering work in this space~\cite{attention-offload-qinghua-2024, pan2024instinfer}. We are actively working toward deploying it over SuperPod.
\end{itemize}

\subsection{Serving at SuperPod-Scale}
\label{sec:flowserve-superpod}

{
\begin{figure}[t]
\begin{center}
\centerline{\includegraphics[width=\linewidth]{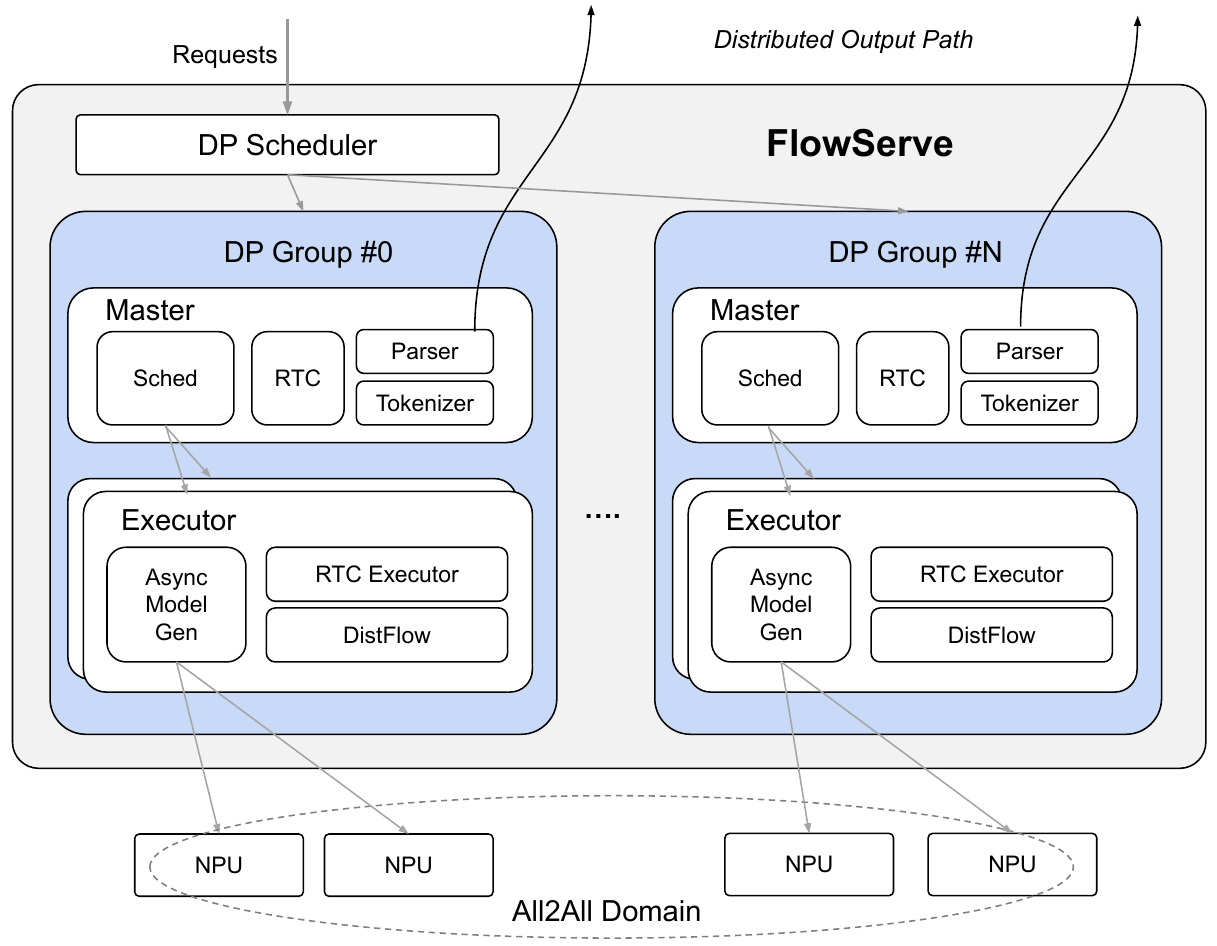}}
\caption{\textbf{Adapting \flowserve\ for SuperPod.} A single \flowserve\ TE can manage up to tens of servers and hundreds of NPU chips.}
\label{fig-flowserve-superpod}
\end{center}
\end{figure}
}

This section outlines the key changes (to \flowserve) that enable it to run at SuperPod scale.
We begin by examining the workload shift that motivates these changes, followed by a description of the corresponding architectural changes.

From a workload perspective, large Mixture-of-Experts (MoE) models---such as DeepSeek-V3/R1~\cite{liu2024deepseek-v3} and Pangu-Ultra-MoE~\cite{pangu-moe-2025}---are a natural fit for SuperPod-scale infrastructure.
First, MoE models require efficient all-to-all communication before and after the expert modules, benefiting significantly from a large-scaled-up domain with high-bandwidth interconnects.
Second, MoE experts achieve high compute efficiency at only large batch sizes. Enabling such batches necessitates data parallelism in the attention layers (DP Attention) instead of traditional tensor parallelism. Under DP Attention, colocating the prefill and decode stages leads to severe interference~\cite{liu2024deepseek-v3,tetriserve-arxiv24}. As a result, effective deployment requires disaggregating prefill and decode (PD disaggregation), which further depends on fast, low-latency interconnects.
The \cloudmatrix\ SuperPod provides a significantly larger scaled-up domain with a global shared memory address space and enables low-latency, high-bandwidth data transfers between any CPU or NPU in the system, making it ideally suited for serving large-scale MoE models.

Motivated by these requirements, we introduce two major architectural changes to scale \flowserve\ to hundreds of NPU chips, as summarized in Figure~\ref{fig-flowserve-superpod}.
First, to avoid scheduling bottlenecks, we partition the system into multiple parallel DP groups. Each group includes a full set of components—its own scheduler, RTC, DistFlow, and executors.
A centralized DP scheduler dispatches requests across DP groups using either a round-robin or greedy strategy. To eliminate bottlenecks on the return path, each DP group independently returns tokens to the frontends.
Second, to support large-scale PD disaggregation, we extend \texttt{DistFlow} to support M:N connections. Specifically, we establish transfer channels between all prefill and decode DP groups.

\section{Distributed Scheduling}
\label{sec-dist-sched}

This section discusses \sysname's distributed LLM request scheduling policies.

\subsection{Overview}
\label{sec-sched-overview}

In \sysname, distributed scheduling for LLM requests faces three new challenges.
The first challenge involves locality and states. 
Before prefix caching was introduced into \flowserve, distributed scheduling was straightforward, as all TEs could be treated as stateless, and scheduling was based solely on load. 
However, with the introduction of prefix caching, TEs have become stateful, and the goal is to schedule requests where KV cache reuse is possible, leading to the need for locality-aware scheduling.
Achieving locality-aware scheduling is non-trivial. 
Selecting the best TE for a given request requires balancing KV cache reuse while avoiding overloading the TE that holds the most cache.
The second challenge comes from disaggregated serving. 
Distributed scheduling was easier before PD-disaggregation was added to \flowserve\ because all TEs were the same. However, with PD disaggregation, choosing the best TE for a request becomes more complex, as it is unclear whether PD-disaggregated TEs or a PD-colocated TE will perform better.
The last challenge arises from the coexistence of both prefix caching and disaggregation in \sysname, requiring any practical algorithm to consider both factors.

\sysname\ uses the following designs in response to these challenges: a locality-aware scheduling algorithm (\S\ref{sec-sched-la}), a PD-aware scheduling algorithm (\S\ref{sec-sched-pda}), and a combined scheduling algorithm (\S\ref{sec-sched-combined}) as listed in Algorithm~\ref{alg:dist-sched}.

\subsection{Locality-aware Scheduling}
\label{sec-sched-la}

In this section, we aim to answer one question: 

\begin{mdframed}[style=dashedbox]
Given a request and a set of TEs with associated KV cache residency information, how can we select the TE that maximizes KV cache reuse to efficiently serve the request?
\end{mdframed}

To address this question, we implement a prompt-tree-based, locality-aware scheduling policy in the \texttt{select\_tes\_prefix\_match} function, as shown in Algorithm~\ref{alg:dist-sched}.
Similar to Preble~\cite{srivatsa2024preble}, SGLang~\cite{sglang}, and MemServe~\cite{hu2024memserve}, the distributed scheduler in JE maintains a global prompt tree for each type of TE, while each TE also maintains a local prompt tree that shares an index with its corresponding global tree. When a request arrives at the JE, it matches the prompt tokens against the global prompt trees and selects the TE that has the longest common prefix, indicating the largest preserved historical KV cache.

\subsection{PD-aware Scheduling}
\label{sec-sched-pda}

{
\begin{figure*}[t]
\begin{center}
\centerline{\includegraphics[width=0.9\linewidth]{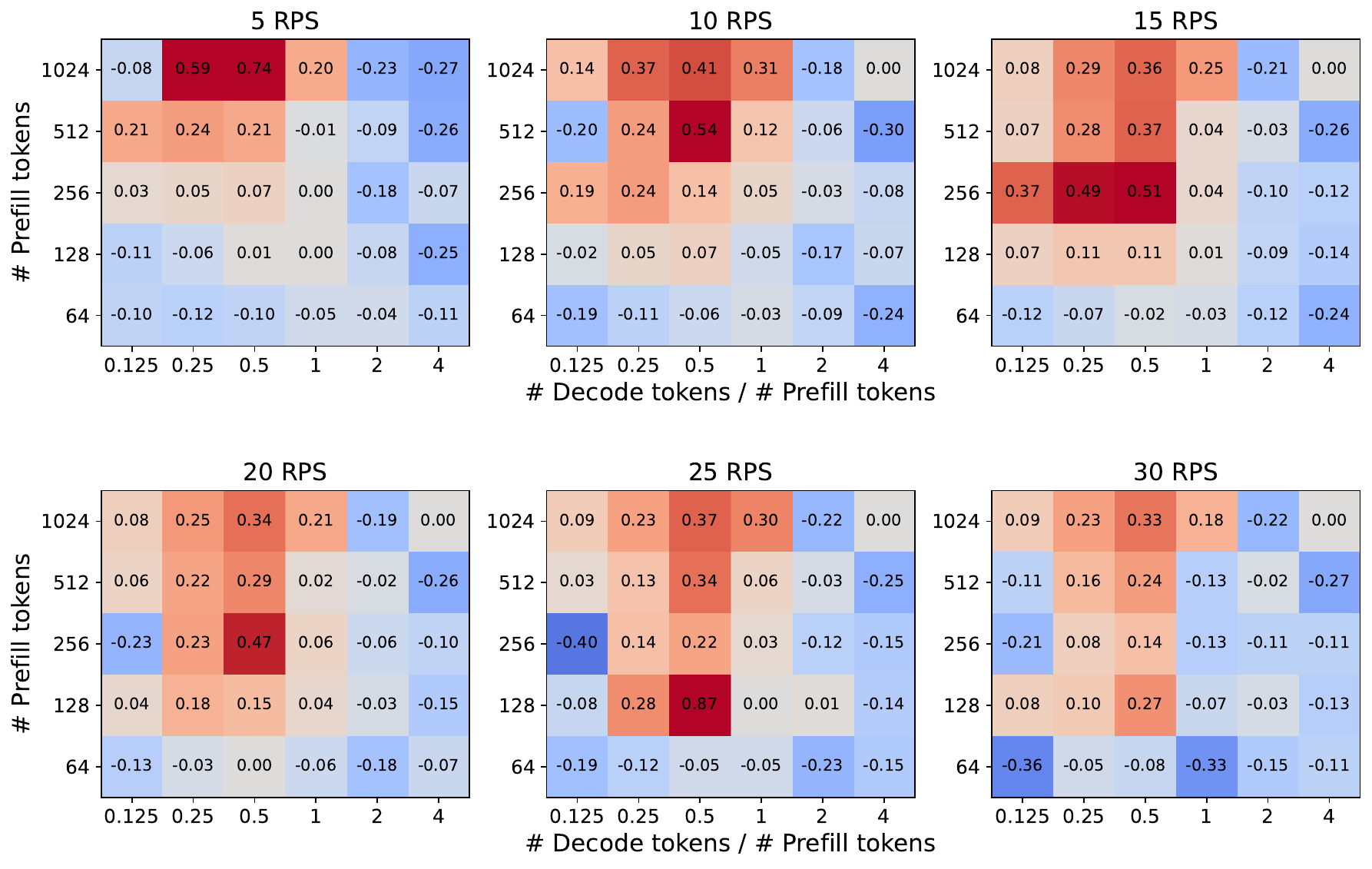}}
\caption{\textbf{Comparing the Performance of PD-disaggregated and PD-colocated (with chunked prefill) using Heatmap.} 
The y-axis represents the prefill length, and the x-axis shows the ratio of decode length to prefill length. 
For each combination of prefill and decode lengths, we execute a batch of identical requests at a fixed RPS on both PD-disaggregated and PD-colocated TEs.
The heatmap cells display the difference in JCT between the two setups, computed as the ratio of JCT for the PD-colocated TE to the PD-disaggregated TE, minus one. 
A positive value indicates better performance of the PD-disaggregated TE, while a negative value suggests that the PD-colocated TE is more efficient. This figure uses a 34B model with TP=4.}

\label{fig-dist-sched-heatmap}
\end{center}
\end{figure*}
}

In this section, we aim to answer one question: 

\begin{mdframed}[style=dashedbox]
Given a request and a set of TEs with both PD-disaggregated and PD-colocated instances, how can we select the most appropriate TE type that best aligns with the characteristics of the request?
\end{mdframed}

We first study the performance comparison between PD-disaggregated and PD-colocated across various dimensions in a heatmap.
We then discuss converting such a heatmap into a practical scheduling algorithm.

\subsubsection{Study}

We first run tests to compare PD-disaggregated and PD-colocated TEs (with chunked prefill).  
Figure~\ref{fig-dist-sched-heatmap} presents the results (see caption for detailed setup).

The heat map shows that PD-disaggregated and PD-colocated TEs divide the request space into distinct regions.
We find three key observations.
First, the PD-disaggregated setup performs better for requests with longer prompts and shorter decode. Additionally, as prefill length increases, its advantage becomes more pronounced for requests with longer decode lengths.
Second, PD-disaggregated TEs provide a larger performance advantage over PD-colocated TEs (dark read) than the reverse (light blue), suggesting that a correct choice of PD-disaggregated TE leads to significant performance gains, while an incorrect choice results in minimal loss.
Finally, the advantage of PD-disaggregated and PD-colocated TEs remains consistent across different RPS values.

\subsubsection{Algorithm}

The preceding study is conducted in a controlled environment, and applying it to a practical scheduling algorithm is challenging due to the dynamic nature of RPS and the uncertainty of decode length at the time of scheduling. There are multiple potential approaches to convert the study into an algorithm.

We propose a simple policy called \texttt{select-tes-PD-heatmap}, described in Algorithm~\ref{alg:dist-sched}.
The policy works as follows.
First, we combine the heat maps across all RPS values through element-wise addition. Given the stability of the heat map, over 80\% of the squares consistently show either positive or negative values across all RPSs, while the remaining 20\% are uncertain.
Second, we predict the decode length for an incoming request using a predict model with 84.9\% accuracy to balance prediction precision and overhead. For more details, please refer to the following discussion section.
Third, we identify the corresponding square on the heat map based on the combined heat map, the prefill length, and the predicted decode length. If the value is positive, we select PD-disaggregated TEs; if negative, we select PD-colocated TEs.

{
\begin{figure}[th]
\begin{center}
\centerline{\includegraphics[width=0.5\textwidth]{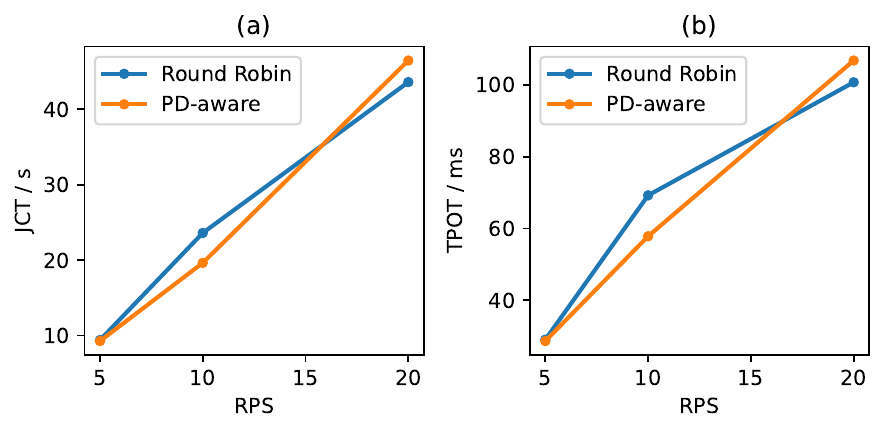}}
\caption{
%Comparison of JCT and TPOT Between Round Robin and PD-Aware Scheduling.
\textbf{Study of Distributed Scheduling Algorithm.}
%
%To demonstrate the advantages of the PD-aware scheduling policy, we design a workload consisting of 800 requests uniformly sampled from the red region and 800 requests uniformly sampled from the blue region of the heat map.
We run a 34B model with TP=4, and report JCT / TPOT.
We run an internal trace sampled from a code-generation service.
%These requests are processed at a fixed RPS using CodeLlama 34B with TP=4.
The cluster consists of four servers with two PD-colocated TEs and a pair of PD-disaggregated TEs (1P1D).
}
\label{fig-eval-profile-pd}
\end{center}
\end{figure}
}

We perform a microbenchmark study shown in Figure~\ref{fig-eval-profile-pd}. We make three key observations.
First, under certain RPS levels (e.g., 10 reqs/s), the PD-aware scheduling policy outperforms RR.
Second, at low RPS levels, the performance of PD-aware matches that of RR. The reason is that under low RPS, interference between prefill and decode operations within PD-colocated TEs remains negligible, and PD-disaggregated TEs do not offer significant advantages.
Third, at very high RPS levels, PD-aware performs worse than RR. The reason is that PD-disaggregated TEs, with the same resources (e.g., two cards), are more prone to overloading. However, even when overloaded, PD-aware scheduling does not exhibit significant performance degradation compared to RR.

\subsubsection{Discussion of the Predict Model} 

Accurately predicting decode lengths with minimal overhead remains a challenging research problem. We adopt the approach proposed in TetriServe~\cite{tetriserve-arxiv24}, which employs a lightweight LLM-based classification model---referred to as the predict model---to classify decode lengths into fixed-size buckets, assuming execution by a specified target LLM.

We predict length ranges rather than exact token counts for two reasons. First, inference parameters such as temperature and top-p~\cite{url-llm-parameter} introduce substantial variability in outputs, making accurate prediction difficult. Second, our goal is to support scheduling decisions, for which approximate length ranges are sufficient (Figure~\ref{fig-dist-sched-heatmap}).

The training process for the predict model consists of the following three steps. First, we construct a prompt-only training dataset using public sources, a target model (e.g., LLaMA 70B), and a predict model (e.g., OPT 125M~\cite{url-hg-opt125}). Second, we query the target model with the prompts to generate responses, which we discretize into fixed-size length buckets to serve as ground truth. Third, we split the dataset into training and evaluation sets, and train the predict model accordingly.

In our experiments, we use a bucket granularity of 128 tokens and achieve 84.9\% accuracy. Since our focus lies in using prediction to inform scheduling, we leave further accuracy improvements to future work.

\subsection{Combined Algorithm}
\label{sec-sched-combined}

\begin{algorithm}
\caption{Distributed Scheduling Policy}\label{alg:dist-sched}

\SetKwData{heatmap}{hm}
\SetKwData{request}{req}
\SetKwData{te}{te}
\SetKwData{tes}{tes}

\SetKwFunction{ds}{dist\_sched}
\SetKwFunction{ilb}{is\_load\_balanced}
\SetKwProg{Fn}{Function}{:}{end}

\SetKwFunction{pda}{PD\_aware}
\SetKwData{pl}{p\_l}
\SetKwData{dl}{d\_l}
\SetKwFunction{gpl}{get\_prefill\_length}
\SetKwFunction{gdl}{get\_decode\_length}
\SetKwFunction{ste}{select\_tes\_PD\_heatmap}

\SetKwFunction{la}{locality\_aware}
\SetKwFunction{spm}{select\_tes\_prefix\_match}

\SetKwFunction{lla}{load\_aware}
\SetKwFunction{sll}{select\_tes\_least\_load}

\KwIn{A new request: \request, a TE group: \tes}
\KwOut{A TE to forward the request to}
\BlankLine

\Fn{\ds{\request, \tes}} {

    \tes $\gets$ \pda(\request, \tes)\;

    \If{\tes.\ilb()}{
        \tes $\gets$ \la(\request, \tes)\;
    }
    \Else {
        \tes $\gets$ \lla(\request, \tes)\;
    }

    \Return \tes\;
}

\Fn{\pda{\request, \tes}}{

    \pl $\gets$ \request.\gpl()\;
    
    \dl $\gets$ \request.\gdl()\;
    
    \tes $\gets$ \tes.\ste(\pl, \dl)\;
    
    \Return \tes\;
}

\Fn{\la{\request, \tes}}{

    \tes $\gets$ \tes.\spm(\request)\;
    
    \Return \tes\;
}

% \Fn{\lla{\request, \tes}}{

%     \tes $\gets$ \tes.\sll()\;
    
%     \Return \tes\;
% }

\end{algorithm}

We propose a combined algorithm that integrates load-aware, locality-aware, and PD-aware scheduling. The scheduling algorithm, outlined in Algorithm~\ref{alg:dist-sched}, relies on three core function calls: \texttt{locality\_aware}, \texttt{PD\_aware}, and \texttt{load\_aware}. These functions progressively refine the TE group, narrowing it down from the entire set to a single TE based on the request's characteristics and the underlying TEs.

The scheduling process proceeds as follows. First, a subgroup of TEs is selected by choosing a specific type, such as PD-colocated or PD-disaggregated TEs, based on the request's length and the heatmap of the TE group (Section~\ref{sec-sched-pda}). Second, once a subgroup of TEs is identified, the selection is further refined. If the load is balanced across the remaining TEs, the algorithm prioritizes a TE with the most reusable KV cache, using tree-based prefix matching (Section~\ref{sec-sched-la}). If the load is unbalanced, the algorithm instead prioritizes a TE with the least load to achieve better load distribution.

\section{Fast Scaling}
\label{sec-scaling}

{
\begin{figure}[t]
\begin{center}
\centerline{\includegraphics[width=0.5\textwidth]{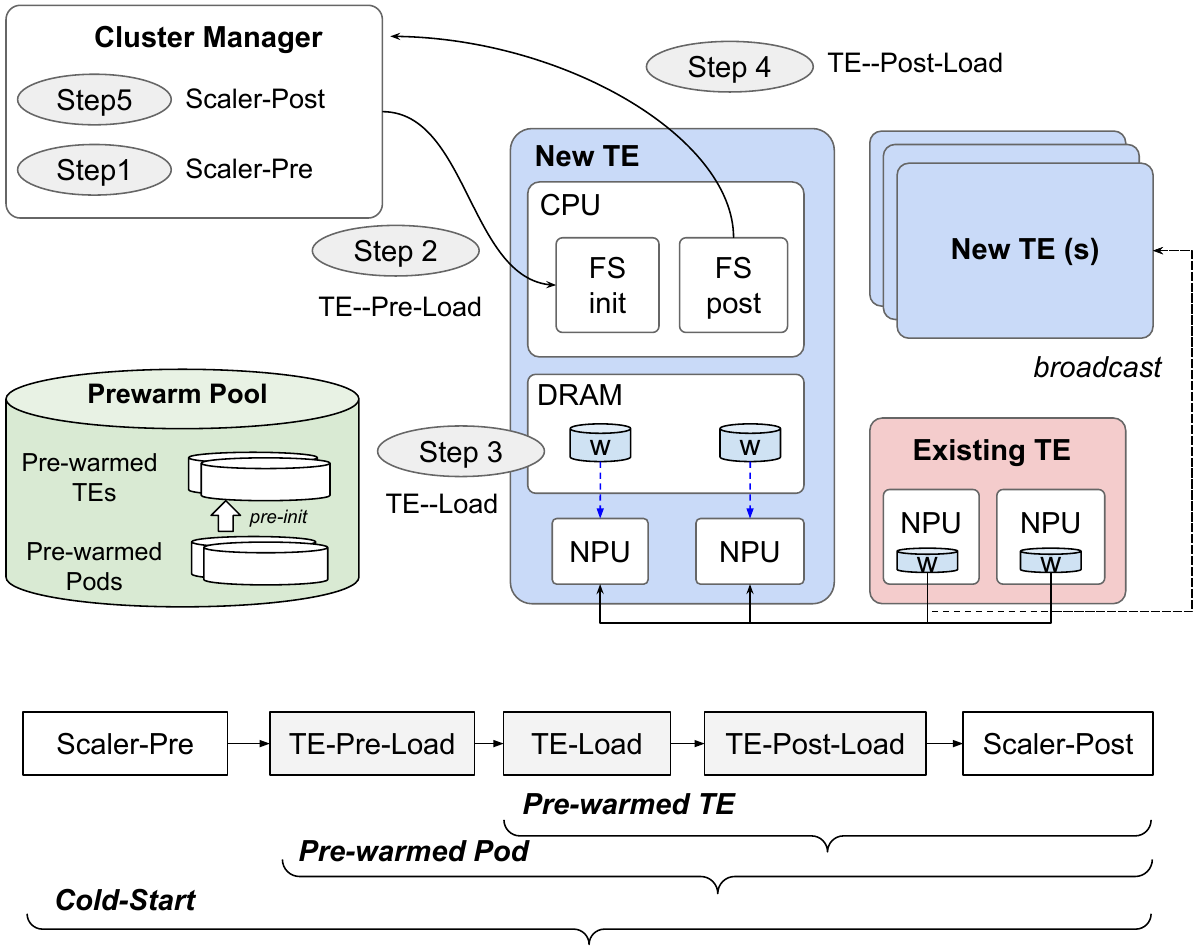}}
\caption{\textbf{\sysname's Scaling Design.} (a) We show TE-Load's two cases: loading from local DRAM (DRAM-hit) and loading from another TE's NPU (NPU-fork). NPU-fork can be either via scale-up or scale-out network link. (c) The bottom shows the timeline for three scaling cases.}
\label{fig-scaling-design}
\end{center}
\end{figure}
}
\begin{table*}[ht]
\caption{\textbf{A Summary of \sysname's End-to-End Scaling Steps, Challenges, and Solutions.}}
\centering
\small % Reduce text size for the entire table
\begin{tabular}{c|l|p{5cm}|p{4cm}|p{3cm}}
\hline
 \textbf{ID} & \textbf{Step} & \textbf{Definition} & \textbf{Major Issues} & \textbf{Our Solutions} \\
\hline

1 & Scaler-Pre & Creating pods to hold the TE. & 1. Resource allocation is slow & 1. Pre-warmed Pods \\
\hline

2 & TE-Pre-Load & Launching the TE w/o model loading & \parbox[l][0.8cm][c]{3cm}{1. Python startup is slow \\ 2. NPU init is slow } & 1. Pre-warmed TEs \\
\hline

3 &  TE-Load & Loading the model onto the NPU & 1. Model weight is large & \parbox[l][1cm][c]{3.5cm}{1. DRAM pre-loading \\ 2. NPU-fork } \\
\hline

4 &  TE-Post-Load & Preparing TE to serve requests & \parbox[l][1cm][c]{3.4cm}{1. Engine warmup is slow \\ 2. Block alloc is slow} & \parbox[l][1.5cm][c]{2.9cm}{1. Offline profiling \\ 2. Async allocation \\3. Dummy req warmup} \\
\hline

5 & Scaler-Post & From TE ready to serve first request  & 1. The update of the global TE list is slow & 1. Proactive pushing  \\
\hline
\end{tabular}

\label{tbl-scaling}
\end{table*}

{
\begin{figure}[h]
\begin{center}
\centerline{\includegraphics[width=0.5\textwidth]{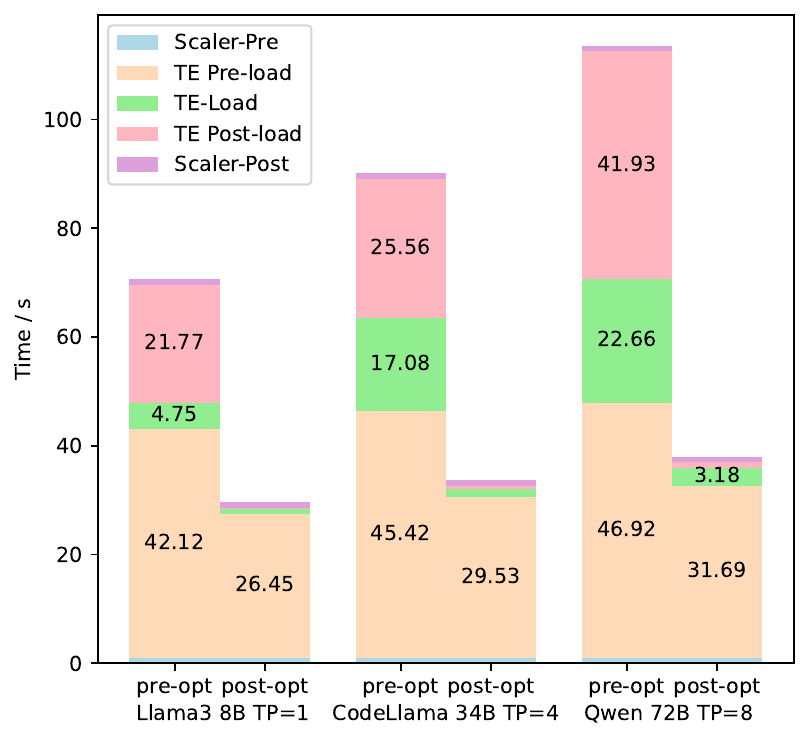}}
\caption{\textbf{Scaling E2E Breakdown}. We present the scaling latency both before and after optimizations. Even after optimization, the \textit{TE-Pre-load} step still remains the dominant factor in scaling time, although this latency can be further reduced through pre-warming (\S\ref{sec:prewarm}).}
\label{fig-scaling-e2e-breakdown}
\end{center}
\end{figure}
}
{
\begin{figure}[t]
\begin{center}
\centerline{\includegraphics[width=\linewidth]{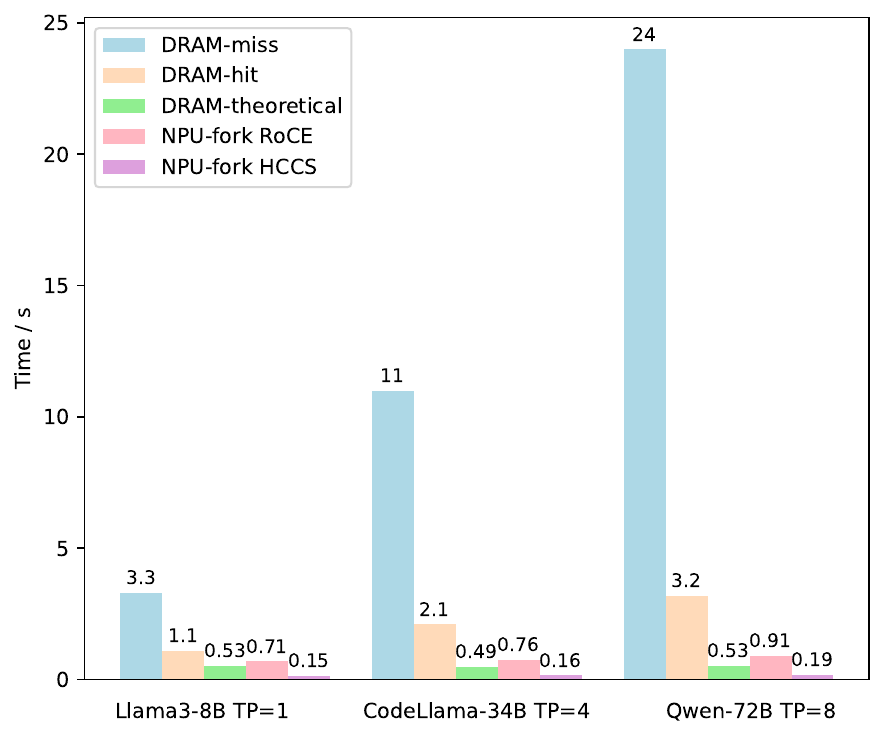}}
\caption{\textbf{TE-Load Study.} DRAM-hit means loading from pre-loaded weights in DRAM; DRAM-miss means 
pre-load miss, thus loading from SSD; DRAM-theoretical is calculated by model weights dividing PCIe bandwidth.
NPU-fork has two different links---HCCS and RoCE (see the main text for the differences).}
%For each model, the orange bar represents the latency of TE-Load from DRAM, while the blue bar indicates the latency of TE-Load from neighboring NPUs.
%The dashed black line shows the theoretical latency, calculated by dividing the model weight size by the theoretical bandwidth from DRAM to NPU or from NPU to NPU.}
\label{fig-scaling-te-load}
\end{center}
\end{figure}
}
{
\begin{figure}[t]
\begin{center}
\centerline{\includegraphics[width=\linewidth]{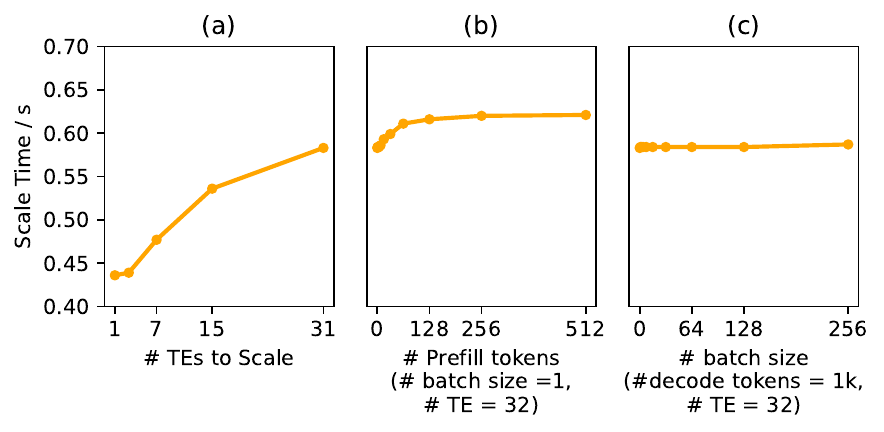}}
\caption{\textbf{Scalability and Sensitivity Study of NPU-fork}. We run NPU-fork over a scaled-up network (HCCS) using Llama3-8B-TP1. (a) Scaling multiple TEs in parallel from one running TE. (b) Time for scaling to 32 TEs when the source TE is prefilling sequences of different lengths. (c) Scaling time when the source TE is decoding different batches of sequences, each with a length of 1k tokens.}
\label{fig-scaling-te-load-npu-hit-sensitivity}
\end{center}
\end{figure}
}

In \sysname, the cluster manager's \autoscaler\ determines when to scale TEs and JEs based on 
metrics such as load or SLO-violation rates.
Fast scaling is critical but particularly challenging for LLM-serving TEs, 
as it requires loading large model weights onto NPUs. 
This section describes our optimizations for fast scaling.

%
%We show our scaling mechanism for LLM serving TE in Figure~\ref{fig-scaling-design} and our optimizations for speed-up scaling in Table~\ref{tbl-scaling}.
%
Figure~\ref{fig-scaling-design} shows the autoscaling workflow, which involves five steps, with challenges and 
solutions summarized in Table~\ref{tbl-scaling} and performance breakdown in Figure~\ref{fig-scaling-e2e-breakdown}. First, in the \textit{Scaler-Pre} step, 
\sysname\ prepares resources and pod environments for the new TEs. For large models, 
a TE may span multiple pods on multiple machines.
Second, the \textit{TE-Pre-Load} step initializes the \flowserve\ instance but does not load the model onto the NPU. 
Third, the \textit{TE-Load} step covers loading model weights onto the NPU. We decouple the second and third steps to allow for 
pre-warmed TEs, facilitating faster scaling (see \S\ref{sec:prewarm}). Fourth, the \textit{TE-Post-Load} step involves engine warm-up and CPU/NPU block allocation. 
Open-source inference engines such as vLLM~\cite{vllm-sosp23} rely on warm-up to profile the allocable HBM size for KV caches. We find this step unnecessary in production, because we can offline profile this data and store it in configurations. To address the slowdown of the first request after removing warm-up, we add a dummy message post-startup.
Finally, in the \textit{Scaler-Post} step, the new TEs are announced to the cluster, 
and JEs direct requests to them.

% The major issues lie in resource allocation, loading library, and synchronous operations.
%

\subsection{Pre-warming Pods and TEs}\label{sec:prewarm}
\sysname\ maintains two levels of pre-warmed resources. 
First, at the pod level, 
\sysname\ reserves a small number of pre-warmed pods with basic environment setup. 
These workload-independent pods are usually managed by the infrastructure layer, 
such as Kubernetes, and can be shared across services to reduce overhead. Second, on these pre-warmed pods, \sysname\ maintains a small pool of pre-warmed TEs to minimize \textit{TE-Pre-Load} time. 
The \textit{TE-Pre-Load} step includes \flowserve’s startup time, which involves loading Python libraries, initializing NPU states, and setting up HCCL cross-NPU interconnections. 
We optimize this step by approximately 35\% for most models using techniques such as late importing and parallel initialization. 
However, as shown in Figure~\ref{fig-scaling-e2e-breakdown}, this step still accounts for the majority of startup time. 
To further reduce the startup time, we move the initialization out of the critical path by incorporating a TE pre-warming mechanism.

Our TE pre-warming design is carried out in two stages. 
First, we make the pre-warmed TEs model-agnostic. 
For example, a TP-8 pre-warmed TE can be adapted to run either a Llama3-70B or a Qwen2-72B model. 
This stage requires carefully distinguishing between model-specific and model-agnostic parameters. 
Second, we make the pre-warmed TEs agnostic to parallelism strategies by recognizing that, regardless of TP/PP/SP configurations, 
all TEs follow a master-SPMD architecture. This stage allows independent pools of pre-warmed SPMD-masters and SPMD-executors, and these pools can be packed on demand.

\subsection{Optimizing Model Loading}
\sysname\ has two model-loading paths: \textit{local loading} via PCIe from local DRAM or SSD, 
and \textit{NPU-fork} using high-speed NPU-to-NPU links from a running TE. While NPU-fork is generally faster,
it has higher hardware requirements and cannot be used during code start (scaling from 0 TE).

\textbf{Local loading with pre-warming.}
We use the safetensors format~\cite{safetensors} for model storage. 
Compared to the native binary format, safetensors reduce serialization costs by storing tensors in contiguous blocks that can be directly mapped into memory and trigger data reads only on page faults. 
Safetensors also simplify pre-loading into the page cache, reducing read amplification. 
Instead of loading the entire model file, each TP process loads only the required partition on demand.

To further optimize TE-load performance, we co-design it with our TE pre-warming mechanism. 
The cluster manager predicts models that are likely to scale and pre-loads them into DRAM page cache using pre-warmed TEs. 
When scaling is triggered, the master prioritizes selecting pre-warmed TEs with the required model already loaded. 
On our hardware, each machine has 1.5TB of DRAM, sufficient for pre-loading 10 70B models or 100 7B models.

Figure~\ref{fig-scaling-te-load} shows TE-load's performance under different cases (NPU-fork, DRAM-hit, and DRAM-miss).
DRAM-hit occurs when the cluster manager correctly predicts the models to scale, while DRAM-miss reflects the opposite scenario. 
We also compute the theoretical model-loading time by dividing the model weights by the PCIe bandwidth.
For DRAM-hit, the difference between measured and theoretical time arises from two factors: PyTorch-model tensor initialization (about 0.3s) and PCIe-bandwidth contention. The latter becomes more significant with larger TP ranks due to shared PCIe links. 
For the three models in Figure~\ref{fig-scaling-te-load}, the weights loaded by each NPU are roughly the same, but local loading time increases with larger TP ranks due to PCIe-link sharing among NPUs.

\textbf{NPU-fork}.
NPU clusters typically feature high-speed NPU-to-NPU links designed for distributed training. 
In \S\ref{sec:flowserve-rtc}, we demonstrate how these links are used for efficient inference. 
We further exploit them in our NPU-fork technique to transfer model weights during fast scaling. 
When NPU-fork is triggered, the master notifies a running \flowserve\ TE to connect to a pre-warmed TE,
with model weights transferred using the DistFlow module (\S\ref{sec:flowserve-distflow}).

Figure~\ref{fig-scaling-te-load} shows the performance of NPU-fork. 
On our Ascend hardware, there are two types of links: HCCS (higher bandwidth, smaller scale) and RoCE (lower bandwidth, scalable to thousands of nodes). All evaluations are conducted on cross-node setups.
Overall, loading with HCCS is significantly faster than with RoCE, indicating that NPU-fork will benefit from the SuperPod architecture. The model-loading time is similar across the three models, as the weights to be loaded per NPU are roughly the same. NPU-fork experiences less bandwidth contention compared to local loading, as NPU-fork uses different physical links.

Figure~\ref{fig-scaling-te-load-npu-hit-sensitivity} illustrates potential degradation when multiple expansions (up to 32) occur concurrently or when prefilling and decoding overlap the expansions. NPU-fork can scale to a large number of TEs by transmitting model weights simultaneously to multiple TEs using the \texttt{broadcast} API in the HCCL collective communication library. We also measure resource-contention sensitivity when the source TE handles prefill and decode requests. Since the NPU has dedicated AICPUs for data transfer, contention is limited.

%\section{Evaluation}

\balance

\section{Discussion}
\label{sec-discussion}

We now discuss the broader applicability of \sysname\ across hardware platforms and workloads.

\textbf{NPU-agnostic vs. NPU-specific features.}
Although \sysname\ is built on Ascend NPU chips at Huawei Cloud, its high-level architecture and core design are largely hardware-agnostic.
For example, the ``NPU-fork'' mechanism used for auto-scaling leverages Ascend’s HCCS interconnect for rapid scaling. However, similar techniques can be applied to other hardware platforms with fast inter-chip links, such as NVIDIA GPUs with NVLink.
Most NPU-specific features involve low-level computation and networking primitives. For instance, in the CloudMatrix384 SuperPod, memory across all NPUs is accessible, allowing data transfers between any two memory locations—a capability not available in regular scaled-out AI servers.

\textbf{Generalization to non-LLMs.}
While the paper primarily focuses on LLMs, most components generalize to other model types, such as embedding models and multimodal-understanding models. In fact, all these models run within the \flowserve\ engine similar to vLLM.
%
%Multimodal models require a separate encoder component with a distinct caching mechanism, which is similar to vLLM and is currently under early-stage development, representing a target for future optimization.

\textbf{Fault-Recovery Mechanism.} In the event of a TE or JE failure, \sysname\ reboots the affected component and redirects requests to redundant counterparts. The reboot process ensures recovery within 5 minutes. RTC maintains soft states, which can be recomputed if lost and are append-only. Therefore, we do not implement complex consistency protocols to avoid unnecessary performance overhead for RTC. However, in production, we have not observed a significant impact on service quality, primarily due to (1) system-level redundancy and (2) the SuperPod architecture, which ensures fast and uniform NPU-to-NPU communication regardless of physical location, enabling seamless substitution of failed NPUs.

% \textbf{mention potential public release of anonymized traces (upon approval).} \hjh{Will we release traces?} No.

% \textbf{additional PD-aware scheduling results (especially under heavier loads) and provide resource usage statistics}. \hjh{This should be included in Section~\ref{sec-dist-sched}}

%\textbf{If time permits, a comparison with other open-source LLM serving systems (e.g., vLLM, FastServe) would be a valuable addition to the evaluation section.} \hjh{We skip this.}
\section{Related Work}
\label{sec-related}

In this section, we review related work in four key areas: serverless infrastructure, serving engines, scheduling, and scaling for large language model (LLM) workloads.

\textbf{Serverless Infrastructure.}
Extensive research has focused on optimizing LLM serving in serverless architectures~\cite{batch, infless, mark, infaas, gillis, spatio-temporal, tetris}. Industry solutions, such as AWS SageMaker and Azure ML~\cite{minio}, deliver cloud-native solutions optimized for deploying, managing, and scaling model inference, tailored specifically to meet enterprise needs. Recent open-source systems, including KServe~\cite{kserve}, AIBrix~\cite{aibrix}, NVIDIA Dynamo~\cite{dynamo}, and LeaderWorkerSet (LWS)~\cite{lws}, offer practical implementations with varying degrees of support for cloud deployment, auto-scaling, multi-model, multi-node execution, and so forth. \sysname\ offers similar functionality but differs in two key aspects. First, while existing systems primarily target GPU clusters, \sysname\ is designed for NPU-based clusters. Second, to the best of our knowledge, \sysname\ is the first publicly described platform that introduces request-job-task abstraction, integrating diverse AI workloads from post-training to serving.

\textbf{Serving}.
Serving is a rapidly evolving field, with many open-source engines such as \flowserve\ emerging. vLLM~\cite{vllm-sosp23} pioneers PagedAttention for higher throughput. SGLang~\cite{sglang} uses RadixAttention for reusing cache. LightLLM~\cite{hu2024lightllm} adopts asynchronous execution for improved throughput.
We believe that all serving engines will eventually adopt similar architectures to optimize AI chip efficiency, with differences mainly in hardware support, features, and programming languages.
Networking-wise, the Mooncake Transfer Engine (MTE)~\cite{qin2024mooncake} is closest to \flowserve's DistFlow. Both systems support multiple backends and efficient threading models, following seminal lines of work~\cite{erpc-nsdi19,tsai2017lite,kalia2016fasst}. The key difference lies in backend implementation: MTE uses RDMA by default, while \flowserve\ uses HCCL's peer-to-peer APIs on regular scaled-out servers and memory-copy primitives on the \cloudmatrix\ SuperPod.

%In industry, Kimi~\cite{qin2024mooncake} proposes a disaggregated architecture that centralizes key-value (KV) cache management to optimize LLM serving.
%

\textbf{Scheduling}.
Efficient scheduling is critical for improving the performance of serving systems~\cite{yu2022orca, sarathi-arvix23, fastserve-arxiv23, muxserve, hu2024memserve, qin2024mooncake, vllm-sosp23, sglang, patel2023splitwise, zhong2024distserve}.
%There are two layers: a local layer and a global layer.
%
For example, at the local layer,
Orca~\cite{yu2022orca} proposes iterative-level scheduling to reduce bubbles.
Sarathi~\cite{sarathi-arvix23} proposes chunked-prefill to overcome suboptimal prefill processing.
FastServe\cite{fastserve-arxiv23} utilizes a multi-level priority feedback queue to minimize JCT.
At the global layer,
MuxServe~\cite{muxserve} formulates a multiplexing problem and proposes a placement algorithm and adaptive batch scheduling strategy to identify optimal colocations in LLM serving.
MemServe~\cite{hu2024memserve} prioritizes locality by directing requests to instances with the highest cache-hit rate.
Our work introduces a novel PD-aware scheduling policy to determine whether a request should be processed by PD-disaggregated or PD-colocated TEs.
Additionally, our approach integrates PD awareness with locality- and load-aware scheduling, providing a comprehensive solution for optimizing resource utilization and throughput across heterogeneous TE configurations.

\textbf{Scaling Optimizations.}
Scaling serving instances dynamically is a major challenge, mainly because these instances are large and continue growing as modern LLMs' size increases.
Recent work on model autoscaling~\cite{pipeswitch, deepplan, spotserve, blizscale-arxiv24} has focused on improving resource use and reducing scaling delays. For example, SpotServe~\cite{spotserve} and Llumnix~\cite{llumnix} speed up scaling by making task migration between instances cheaper. BlitzScale~\cite{blizscale-arxiv24} improves throughput when loading parameters, reduces delays in handling burst requests, and improves the overall service.
Our NPU-fork mechanism is similar to BlitzScale~\cite{blizscale-arxiv24} but differs in the underlying network fabrics.

\section{Conclusion}

We have presented \sysname, a serverless AI platform developed at Huawei Cloud. We have described its serverless abstraction and infrastructure, which enables efficient management of AI workloads. We have also outlined the architecture of \flowserve, our in-house serving engine, detailing its design principles and core components. Additionally, we have explored distributed scheduling policies across a heterogeneous pool of serving instances. Finally, we have provided an end-to-end analysis of fast scaling, highlighting the techniques that allow \sysname\ to quickly adjust to fluctuating workloads.

\section*{Acknowledgment}
This work was partially supported by National Natural Science Foundation of China under Grant No. 92464301. We would also like to thank our shepherd Yue Cheng and other anonymous reviewers for their insightful comments and suggestions, which greatly help improve the quality of this paper. 

\clearpage

\bibliographystyle{plain}
\bibliography{paper}

\begin{thebibliography}{10}

\bibitem{sarathi-arvix23}
Amey Agrawal, Ashish Panwar, Jayashree Mohan, Nipun Kwatra, Bhargav~S Gulavani, and Ramachandran Ramjee.
\newblock {SARATHI}: Efficient {LLM} inference by piggybacking decodes with chunked prefills.
\newblock {\em CoRR}, 2023.

\bibitem{batch}
Ahsan Ali, Riccardo Pinciroli, Feng Yan, and Evgenia Smirni.
\newblock Batch: machine learning inference serving on serverless platforms with adaptive batching.
\newblock In {\em Proceedings of the International Conference for High Performance Computing, Networking, Storage and Analysis}, pages 1--15, 2020.

\bibitem{url-llm-parameter}
{AWS Bedrock}.
\newblock {AWS Bedrock}.
\newblock \url{https://docs.aws.amazon.com/bedrock/latest/userguide/inference-parameters.html}.

\bibitem{pipeswitch}
Zhihao Bai, Zhen Zhang, Yibo Zhu, and Xin Jin.
\newblock {PipeSwitch}: Fast pipelined context switching for deep learning applications.
\newblock In {\em Proceedings of the 14th {USENIX} Symposium on Operating Systems Design and Implementation}, pages 499--514, 2020.

\bibitem{pathways-mlsys22}
Paul Barham, Aakanksha Chowdhery, Jeff Dean, Sanjay Ghemawat, Steven Hand, Daniel Hurt, Michael Isard, Hyeontaek Lim, Ruoming Pang, Sudip Roy, Brennan Saeta, Parker Schuh, Ryan Sepassi, Laurent Shafey, Chandu Thekkath, and Yonghui Wu.
\newblock Pathways: Asynchronous distributed dataflow for {ML}.
\newblock In {\em Proceedings of the Machine Learning and Systems}, pages 430--449, 2022.

\bibitem{aibrix}
{ByteDance}.
\newblock {AIBrix}.
\newblock \url{https://github.com/vllm-project/aibrix}.

\bibitem{attention-offload-qinghua-2024}
Shaoyuan Chen, Yutong Lin, Mingxing Zhang, and Yongwei Wu.
\newblock Efficient and economic large language model inference with attention offloading.
\newblock {\em CoRR}, 2024.

\bibitem{spatio-temporal}
Seungbeom Choi, Sunho Lee, Yeonjae Kim, Jongse Park, Youngjin Kwon, and Jaehyuk Huh.
\newblock Serving heterogeneous machine learning models on multi-{GPU} servers with spatio-temporal sharing.
\newblock In {\em Proceedings of the 2022 {USENIX} Annual Technical Conference}, pages 199--216, 2022.

\bibitem{muxserve}
Jiangfei Duan, Runyu Lu, Haojie Duanmu, Xiuhong Li, Xingcheng Zhang, Dahua Lin, Ion Stoica, and Hao Zhang.
\newblock {MuxServe}: Flexible spatial-temporal multiplexing for multiple {LLM} serving.
\newblock In {\em Proceedings of the 41st International Conference on Machine Learning}, pages 11905--11917, 2024.

\bibitem{fu2024serverlessllm}
Yao Fu, Leyang Xue, Yeqi Huang, Andrei{-}Octavian Brabete, Dmitrii Ustiugov, Yuvraj Patel, and Luo Mai.
\newblock {ServerlessLLM}: Low-latency serverless inference for large language models.
\newblock In {\em Proceedings of the 18th {USENIX} Symposium on Operating Systems Design and Implementation}, pages 135--153, 2024.

\bibitem{hu2024memserve}
Cunchen Hu, Heyang Huang, Junhao Hu, Jiang Xu, Xusheng Chen, Tao Xie, Chenxi Wang, Sa~Wang, Yungang Bao, Ninghui Sun, et~al.
\newblock Memserve: Context caching for disaggregated {LLM} serving with elastic memory pool.
\newblock {\em CoRR}, 2024.

\bibitem{tetriserve-arxiv24}
Cunchen Hu, Heyang Huang, Liangliang Xu, Xusheng Chen, Jiang Xu, Shuang Chen, Hao Feng, Chenxi Wang, Sa~Wang, Yungang Bao, et~al.
\newblock Inference without interference: Disaggregate {LLM} inference for mixed downstream workloads.
\newblock {\em CoRR}, 2024.

\bibitem{hu2024lightllm}
Jiawei Hu, Hong Jia, Mahbub Hassan, Lina Yao, Brano Kusy, and Wen Hu.
\newblock {LightLLM}: {A} versatile large language model for predictive light sensing.
\newblock In {\em Proceedings of the 23rd {ACM} Conference on Embedded Networked Sensor Systems}, pages 158--171, 2025.

\bibitem{hu2025epic}
Junhao Hu, Wenrui Huang, Haoyi Wang, Weidong Wang, Tiancheng Hu, Qin Zhang, Hao Feng, Xusheng Chen, Yizhou Shan, and Tao Xie.
\newblock {EPIC:} efficient position-independent caching for serving large language models.
\newblock In {\em Proceedings of the 42nd International Conference on Machine Learning}, 2025.

\bibitem{hu2025raas}
Junhao Hu, Wenrui Huang, Weidong Wang, Zhenwen Li, Tiancheng Hu, Zhixia Liu, Xusheng Chen, Tao Xie, and Yizhou Shan.
\newblock {RaaS}: Reasoning-aware attention sparsity for efficient llm reasoning.
\newblock In {\em Proceedings of the 63rd Annual Meeting of the Association for Computational Linguistics}, 2024.

\bibitem{mindie-atb}
Huawei.
\newblock Mindie atb models.
\newblock \url{https://www.hiascend.com/document/detail/zh/mindie/10RC2/mindiellm/llmdev/mindie_llm0004.html}, 2025.
\newblock Accessed: 2025-01-14.

\bibitem{url-hg-opt125}
{HugginFace}.
\newblock {HugginFace}.
\newblock \url{https://huggingface.co/docs/transformers/model_doc/opt#transformers.OPTForSequenceClassification}.

\bibitem{deepplan}
Jinwoo Jeong, Seungsu Baek, and Jeongseob Ahn.
\newblock Fast and efficient model serving using multi-{GPUs} with direct-host-access.
\newblock In {\em Proceedings of the 18th European Conference on Computer Systems}, pages 249--265, 2023.

\bibitem{kalia2016fasst}
Anuj Kalia, Michael Kaminsky, and David~G. Andersen.
\newblock {FaSST}: Fast, scalable and simple distributed transactions with two-sided {(RDMA)} datagram {RPC}s.
\newblock In {\em Proceedings of the 12th {USENIX} Symposium on Operating Systems Design and Implementation}, pages 185--201, 2016.

\bibitem{erpc-nsdi19}
Anuj Kalia, Michael Kaminsky, and David~G. Andersen.
\newblock Datacenter {RPC}s can be general and fast.
\newblock In {\em Proceedings of the 16th {USENIX} Symposium on Networked Systems Design and Implementation}, pages 1--16, 2019.

\bibitem{kserve}
{KServe}.
\newblock {KServe}.
\newblock \url{https://github.com/kserve/kserve}.

\bibitem{vllm-sosp23}
Woosuk Kwon, Zhuohan Li, Siyuan Zhuang, Ying Sheng, Lianmin Zheng, Cody~Hao Yu, Joseph Gonzalez, Hao Zhang, and Ion Stoica.
\newblock Efficient memory management for large language model serving with {PagedAttention}.
\newblock In {\em Proceedings of the 29th Symposium on Operating Systems Principles}, pages 611--626, 2023.

\bibitem{tetris}
Jie Li, Laiping Zhao, Yanan Yang, Kunlin Zhan, and Keqiu Li.
\newblock Tetris: Memory-efficient serverless inference through tensor sharing.
\newblock In {\em Proceedings of the 2022 {USENIX} Annual Technical Conference}, pages 572--488, 2022.

\bibitem{ascend-npu-hotchips19}
Heng Liao, Jiajin Tu, Jing Xia, and Xiping Zhou.
\newblock {DaVinci}: {A} scalable architecture for neural network computing.
\newblock In {\em Proceedings of the 2019 {IEEE} Hot Chips Symposium}, pages 1--44, 2019.

\bibitem{liu2024deepseek-v3}
Aixin Liu, Bei Feng, Bing Xue, Bingxuan Wang, Bochao Wu, Chengda Lu, Chenggang Zhao, Chengqi Deng, Chenyu Zhang, Chong Ruan, et~al.
\newblock {DeepSeek}-v3 technical report.
\newblock {\em CoRR}, 2024.

\bibitem{lws}
{LWS}.
\newblock {LWS}.
\newblock \url{https://github.com/kubernetes-sigs/lws}.

\bibitem{spotserve}
Xupeng Miao, Chunan Shi, Jiangfei Duan, Xiaoli Xi, Dahua Lin, Bin Cui, and Zhihao Jia.
\newblock {SpotServe}: Serving generative large language models on preemptible instances.
\newblock In {\em Proceedings of the 29th {ACM} International Conference on Architectural Support for Programming Languages and Operating Systems}, pages 1112--1127, 2024.

\bibitem{minio}
{MinIO}.
\newblock {MinIO}.
\newblock \url{https://min.io}.

\bibitem{monga2021birds}
Sumit~Kumar Monga, Sanidhya Kashyap, and Changwoo Min.
\newblock Birds of a feather flock together: Scaling {RDMA} {RPC}s with flock.
\newblock In {\em Proceedings of the {ACM} {SIGOPS} 28th Symposium on Operating Systems Principles}, pages 212--227, 2021.

\bibitem{dynamo}
{NVIDIA}.
\newblock {Dynamo}.
\newblock \url{https://github.com/ai-dynamo/dynamo}.

\bibitem{openai-url}
OpenAI.
\newblock Openai, 2025.
\newblock Accessed: 2025-01-14.

\bibitem{pan2024instinfer}
Xiurui Pan, Endian Li, Qiao Li, Shengwen Liang, Yizhou Shan, Ke~Zhou, Yingwei Luo, Xiaolin Wang, and Jie Zhang.
\newblock Instinfer: In-storage attention offloading for cost-effective long-context llm inference.
\newblock {\em CoRR}, 2024.

\bibitem{patel2023splitwise}
Pratyush Patel, Esha Choukse, Chaojie Zhang, Aashaka Shah, {\'{I}}{\~{n}}igo Goiri, Saeed Maleki, and Ricardo Bianchini.
\newblock Splitwise: Efficient generative {LLM} inference using phase splitting.
\newblock In {\em Proceedings of the 51st {ACM/IEEE} Annual International Symposium on Computer Architecture}, pages 118--132, 2024.

\bibitem{qin2024mooncake}
Ruoyu Qin, Zheming Li, Weiran He, Jialei Cui, Feng Ren, Mingxing Zhang, Yongwei Wu, Weimin Zheng, and Xinran Xu.
\newblock Mooncake: Trading more storage for less computation - {A} {KVCache}-centric architecture for serving {LLM} chatbot.
\newblock In {\em Proceedings of the 23rd {USENIX} Conference on File and Storage Technologies}, pages 155--170, 2025.

\bibitem{infaas}
Francisco Romero, Qian Li, Neeraja~J. Yadwadkar, and Christos Kozyrakis.
\newblock {INFaaS}: Automated model-less inference serving.
\newblock In {\em Proceedings of the 2021 {USENIX} Annual Technical Conference}, pages 397--411, 2021.

\bibitem{safetensors}
The safetensors contributors.
\newblock Safetensors: A safe and efficient format for tensor serialization, 2023.
\newblock Accessed: 2025-01-14.

\bibitem{url-cloudmatrix}
{SemiAnalysis}.
\newblock {Huawei AI CloudMatrix384}.
\newblock \url{https://semianalysis.com/2025/04/16/huawei-ai-cloudmatrix-384-chinas-answer-to-nvidia-gb200-nvl72/}.

\bibitem{fully-disagg-apsys22}
Yizhou Shan, Will Lin, Zhiyuan Guo, and Yiying Zhang.
\newblock Towards a fully disaggregated and programmable data center.
\newblock In {\em Proceedings of the 13th ACM SIGOPS Asia-Pacific Workshop on Systems}, 2022.

\bibitem{srivatsa2024preble}
Vikranth Srivatsa, Zijian He, Reyna Abhyankar, Dongming Li, and Yiying Zhang.
\newblock Preble: Efficient distributed prompt scheduling for {LLM} serving.
\newblock In {\em Proceedings of the 13th International Conference on Learning Representations}, 2025.

\bibitem{llumnix}
Biao Sun, Ziming Huang, Hanyu Zhao, Wencong Xiao, Xinyi Zhang, Yong Li, and Wei Lin.
\newblock Llumnix: Dynamic scheduling for large language model serving.
\newblock In {\em Proceedings of the 18th {USENIX} Symposium on Operating Systems Design and Implementation}, pages 173--191, 2024.

\bibitem{pangu-moe-2025}
Yehui Tang, Yichun Yin, Yaoyuan Wang, Hang Zhou, Yu~Pan, Wei Guo, Ziyang Zhang, Miao Rang, Fangcheng Liu, Naifu Zhang, Binghan Li, Yonghan Dong, Xiaojun Meng, Yasheng Wang, Dong Li, Yin Li, Dandan Tu, Can Chen, Youliang Yan, Fisher Yu, Ruiming Tang, Yunhe Wang, Botian Huang, Bo~Wang, Boxiao Liu, Changzheng Zhang, Da~Kuang, Fei Liu, Gang Huang, Jiansheng Wei, Jiarui Qin, Jie Ran, Jinpeng Li, Jun Zhao, Liang Dai, Lin Li, Liqun Deng, Peifeng Qin, Pengyuan Zeng, Qiang Gu, Shaohua Tang, Shengjun Cheng, Tao Gao, Tao Yu, Tianshu Li, Tianyu Bi, Wei He, Weikai Mao, Wenyong Huang, Wulong Liu, Xiabing Li, Xianzhi Yu, Xueyu Wu, Xu~He, Yangkai Du, Yan Xu, Ye~Tian, Yimeng Wu, Yongbing Huang, Yong Tian, Yong Zhu, Yue Li, Yufei Wang, Yuhang Gai, Yujun Li, Yu~Luo, Yunsheng Ni, Yusen Sun, Zelin Chen, Zhe Liu, Zhicheng Liu, Zhipeng Tu, Zilin Ding, and Zongyuan Zhan.
\newblock Pangu ultra {MoE}: How to train your big {MoE} on {Ascend} {NPU}s.
\newblock {\em CoRR}, 2025.

\bibitem{tsai2017lite}
Shin{-}Yeh Tsai and Yiying Zhang.
\newblock {LITE} kernel {RDMA} support for datacenter applications.
\newblock In {\em Proceedings of the 26th Symposium on Operating Systems Principles}, pages 306--324, 2017.

\bibitem{fastserve-arxiv23}
Bingyang Wu, Yinmin Zhong, Zili Zhang, Gang Huang, Xuanzhe Liu, and Xin Jin.
\newblock Fast distributed inference serving for large language models.
\newblock {\em CoRR}, 2023.

\bibitem{infless}
Yanan Yang, Laiping Zhao, Yiming Li, Huanyu Zhang, Jie Li, Mingyang Zhao, Xingzhen Chen, and Keqiu Li.
\newblock {INFless}: a native serverless system for low-latency, high-throughput inference.
\newblock In {\em Proceedings of the 27th {ACM} International Conference on Architectural Support for Programming Languages and Operating Systems}, pages 768--781, 2022.

\bibitem{yao2024cacheblend}
Jiayi Yao, Hanchen Li, Yuhan Liu, Siddhant Ray, Yihua Cheng, Qizheng Zhang, Kuntai Du, Shan Lu, and Junchen Jiang.
\newblock {CacheBlend}: Fast large language model serving for {RAG} with cached knowledge fusion.
\newblock In {\em Proceedings of the 20th European Conference on Computer Systems}, pages 94--109, 2025.

\bibitem{yu2022orca}
Gyeong{-}In Yu, Joo~Seong Jeong, Geon{-}Woo Kim, Soojeong Kim, and Byung{-}Gon Chun.
\newblock Orca: {A} distributed serving system for transformer-based generative models.
\newblock In {\em Proceedings of the 16th {USENIX} Symposium on Operating Systems Design and Implementation}, pages 521--538, 2022.

\bibitem{gillis}
Minchen Yu, Zhifeng Jiang, Hok~Chun Ng, Wei Wang, Ruichuan Chen, and Bo~Li.
\newblock Gillis: Serving large neural networks in serverless functions with automatic model partitioning.
\newblock In {\em Proceedings of the 41st {IEEE} International Conference on Distributed Computing Systems}, pages 138--148, 2021.

\bibitem{mark}
Chengliang Zhang, Minchen Yu, Wei Wang, and Feng Yan.
\newblock {MArk}: Exploiting cloud services for cost-effective, {SLO}-aware machine learning inference serving.
\newblock In {\em Proceedings of the 2019 {USENIX} Annual Technical Conference}, pages 1049--1062, 2019.

\bibitem{blizscale-arxiv24}
Dingyan Zhang, Haotian Wang, Yang Liu, Xingda Wei, Yizhou Shan, Rong Chen, and Haibo Chen.
\newblock Fast and live model auto scaling with {O}(1) host caching.
\newblock {\em CoRR}, 2024.

\bibitem{sglang}
Lianmin Zheng, Liangsheng Yin, Zhiqiang Xie, Jeff Huang, Chuyue Sun, Cody\_Hao Yu, Shiyi Cao, Christos Kozyrakis, Ion Stoica, Joseph~E Gonzalez, et~al.
\newblock Efficiently programming large language models using {SGLang}.
\newblock {\em CoRR}, 2023.

\bibitem{zhong2024distserve}
Yinmin Zhong, Shengyu Liu, Junda Chen, Jianbo Hu, Yibo Zhu, Xuanzhe Liu, Xin Jin, and Hao Zhang.
\newblock {DistServe}: Disaggregating prefill and decoding for goodput-optimized large language model serving.
\newblock In {\em Proceedings of the 18th {USENIX} Symposium on Operating Systems Design and Implementation}, pages 193--210, 2024.

\bibitem{zhu2024nanoflow}
Kan Zhu, Yilong Zhao, Liangyu Zhao, Gefei Zuo, Yile Gu, Dedong Xie, Yufei Gao, Qinyu Xu, Tian Tang, Zihao Ye, et~al.
\newblock Nanoflow: Towards optimal large language model serving throughput.
\newblock {\em CoRR}, 2024.

\end{thebibliography}

\end{document}